\documentclass{article}
\usepackage{helvet,times}
\usepackage{bm,textcomp}

\usepackage{amsmath,amstext,amsxtra,amsfonts,amssymb}
\usepackage{times}
\usepackage{fullpage}
\usepackage{graphicx}
\usepackage{epstopdf}
\usepackage{listings}
\usepackage{courier}
\usepackage{color,soul}

\usepackage{float}

\usepackage[outercaption]{sidecap}
\usepackage[round,numbers,sort&compress]{natbib}

\newcommand\argmax{\operatornamewithlimits{\text{argmax}}}
\newcommand\argmin{\operatornamewithlimits{\text{argmin}}}

\newcommand\M{\mathcal{M}}
\newcommand\T{\mathcal{T}}
\newcommand\N{\,\mathcal{N}}
\renewcommand\L{\mathcal{L}}
\renewcommand\d[1]{\mathrm{d}{#1}\,}
\newcommand\eps{\varepsilon}
\newcommand\Exp[1]{\exp{\left[#1\right]}}
\renewcommand\P[2]{P\!\left({#1}\,|\,{#2}\right)}
\newcommand\rhovec{\bm{\rho}}
\newcommand\Jump{S}
\newcommand\jump{s}
\newcommand\Xhat{\widehat{X}}
\newcommand\intII{\int\!}
\newcommand\intRNpone{\int_{\mathbb{R}^{N+1}}\!\!}
\newcommand\intRN{\int_{\mathbb{R}^N}\!\!}
\newcommand\intRNI{\int_{\mathbb{R}^{N+1}}\!\!}

\newcommand\transp{\intercal}
\newcommand\E[1]{\left\langle{#1}\right\rangle}
\newcommand\logdet[1]{\ln{\det{#1}}}

\newcommand\bigO[1]{\mathcal{O}\left({#1}\right)}

\begin{document}

\setcounter{page}{1} 

\title{Estimation of the Diffusion Constant from Intermittent Trajectories with Variable Position Uncertainties}

\author{Peter Relich, \, Mark Olah, \, Patrick Cutler, \, Keith Lidke}


\maketitle 

\begin{abstract}
The movement of a particle described by Brownian motion is quantified by a single parameter, $D$, the diffusion constant.  The estimation of $D$ from a discrete sequence of noisy observations is a fundamental problem in biological single particle tracking experiments since it can report on the environment and/or the state of the particle itself via hydrodynamic radius.  Here we present a method to estimate $D$ that takes into account several effects that occur in practice, that are important for correct estimation of $D$, and that have hitherto not been combined together for estimation of $D$.  These effects are motion blur from finite integration time of the camera, intermittent trajectories, and time-dependent localization uncertainty.  Our estimation procedure, a maximum likelihood estimation, follows directly from the likelihood expression for a discretely observed Brownian trajectory that explicitly includes these effects.  The manuscript begins with the formulation of the likelihood expression and then presents three methods to find the exact solution.  Each method has its own advantages in either computational robustness, theoretical insight, or the estimation of hidden variables.  We then compare our estimator to previously published estimators using a squared log loss function to demonstrate the benefit of including these effects.
\end{abstract}

\newpage

\section{Introduction}
Single Particle Tracking (SPT) is a method to observe and classify the motion of individual particles as \textit{trajectories}: estimates of a particle's position in a sequence of discrete measurement times.  In the field of biological microscopy, SPT has been used for finding and analyzing protein motion in heterogeneous environments like the cellular membrane~\cite{Saxton1997a, Saxton2010} and cytoplasm~\cite{sanamrad2014single,calderon2013quantifying}. The SPT trajectory information can be used to resolve variations in the individual motion of molecules that would otherwise be lost in ensemble imaging techniques.

In the analysis of trajectories, the pure Brownian motion model is often the first model used to describe a trajectory in the absence of prior information about the movement.  The behavior of a single particle dominated by Brownian motion can be described by a normal distribution with the variance term proportional to a single physical scale parameter $D$, the diffusion constant; which makes Brownian motion the simplest model for describing stochastic motion. More complicated behavior could potentially be modeled as Brownian motion with discrete changes in the diffusion constant that could be identified with change point analysis~\cite{monnier2012bayesian}.  Therefore, the estimation of the diffusion constant of a particle from discrete, noisy, and possibly short particle trajectories is a fundamental problem in single particle tracking.

In this manuscript, we focus on the likelihood distribution of $D$.  We present a maximum-likelihood-based approach for estimating the diffusion constant of a particle given an SPT trajectory that includes the individual localization error for each position in the trajectory, the time of the observation, and the camera integration time.  Our approach is based on a direct solution to the likelihood equation for the observation of a particular trajectory.  The need for such an estimation procedure has evolved out of the rapid progress that has been made in SPT analysis techniques over the last few years~\cite{jaqaman2008robust, serge2008dynamic, chenouard2014objective, mont2014new, yoon2008bayesian}.  In particular, some emitter localization techniques can not only accurately resolve the location of an emitter to tens of nanometers, but can also reliably estimate the localization error~\cite{Smith2010b}.  Because the signal to noise ratio of a particle can vary significantly from frame to frame in an image sequence (e.g. from varying background, or photobleaching of the probe), the localization error reported for each observation in a trajectory can also vary significantly from frame to frame.  We have therefore developed an estimator that takes into account this information.
\subsection{Background and Related Work}
Historically, one of the primary techniques for estimating the diffusion constant from trajectories relied on a linear regression of the mean-squared-displacement (MSD) of the tracked particle coordinates as a function of time lag~\cite{Qian1991a}.  In the absence of measurement errors, the observed MSD for pure Brownian motion scales linearly with time lag and intersects at the origin, allowing the direct recovery of the diffusion constant from a linear regression on the well sampled data points.  It has been shown that a regression of the MSD with an offset parameter can be interpreted to account for the cumulative effects of static~\cite{martin2002apparent} and dynamic measurement errors~\cite{Savin2005}.  If the MSD is built using the same data  points for multiple time lags, the correlation between MSD values must also be taken into account in the regression ~\cite{Qian1991a,Michalet2010,Michalet2012}. Although it seems theoretically possible to include individual localization error into the MSD regression, to date this has not been described.

A separate line of work has focused on maximum likelihood approaches to the estimation procedure.  A maximum likelihood estimator works by finding the maximum of a likelihood function $\L(D)=\P{O}{D}$ that gives the probability of observing a particular trajectory $O$, given a diffusion constant $D$.  Ideally this probability should incorporate both the variable localization errors of the trajectory and effect of motion-blur.  The \textit{motion-blur} effect arises from the fact that each localization is performed on data that is acquired over some non-zero exposure time.  Typically camera sensors integrate the signal over the exposure time resulting in a blurring of the particle image. This blurring has important numerical effects on the likelihood function~\cite{Montiel2006}.  A specific solution to the likelihood function has been accurately derived that incorporates the effects of motion-blur but with the caveat that only a single global localization error estimate is used as an input or estimated~\cite{Berglund2010, Michalet2012}.  This estimator is a more robust alternative to the MSD-based estimators because it can implement all trajectory information without incurring systematic error when the data is not well conditioned for a linear regression.  Subsequent work has extended this approach to deal with non-uniformly spaced or intermittent trajectories~\cite{Shuang2013}, however the particular implementation in ~\cite{Shuang2013} did not account for motion blur.  Maximum likelihood estimators are not the only class of diffusion estimators that have evolved recently; continued development on displacement-based estimators has resulted in an estimator that incorporates the effects of covariances between sequentially observed displacements~\cite{vestergaard2014optimal}.

In this work we provide a generalized solution to the likelihood function, incorporating variable localization errors and variable displacement periods, which results in an improvement in estimation accuracy for short trajectories, trajectories with large variations in localization accuracy, and trajectories with intermittently spaced measurements.  In Sec.~\ref{sec:theory} we formulate the diffusion likelihood function to directly incorporate the effects of motion-blur, variable localization errors, and intermittent or non-uniformly spaced observations in time.  We present three independent solutions to this likelihood function.  The first derivation, the \textit{recursive method} (Sec.~\ref{sec:theory-recursive}), is a sequential integration of the nuisance parameters and provides the fastest numerical implementation.  The second derivation, the \textit{Laplace method} (Sec.~\ref{sec:theory-laplace}), utilizes a second order Taylor expansion to express the likelihood as a multivariate Gaussian in the basis of integration.  The Laplace method additionally returns the maximum likelihood values of the true positions given a sampled $D$.  The third derivation, the \textit{Markov method} (Sec.~\ref{sec:theory-markov}), calculates the characteristic function in order to express the likelihood in the basis of displacements.  The \textit{Markov method} allows us to verify that the generalized form of the expression derived in \cite{Berglund2010} is the same distribution as the expressions derived in this manuscript.  The \textit{Markov method} was also instrumental in determining the coefficients necessary to reduce the computational complexity of all the methods (Sec.~\ref{sec:simplerforms}).  Each of these derivations leads to an independent, numerically accurate computational algorithm for estimating the likelihood of $D$ (Sec.~\ref{sec:implementation}), making full use of all the information contained in a noisy trajectory.  The resulting likelihood calculation allows for robust computations in specific problems, such as a maximum likelihood estimator, maximum a posteriori estimate, or change point analysis.  We compare the results of our maximum likelihood estimator (MLE) to the current state of the art estimation software~\cite{Michalet2012} with the squared log loss function and demonstrate that the additional information provided from the localization errors allows for better estimates of $D$ with trajectories parameterized by any non-constant, but otherwise arbitrary distribution of localization variances.

\section{Theory}
\label{sec:theory}
If a diffusing particle is accurately and exactly observed at a discrete sequence of $N+1$ positions $\mathbf{X}=\{\mathbf{x}_i\}^{N+1}_{i=1}$ at times $t_i$, then $\P{\mathbf{X}}{D}$, the probability of sequence $\mathbf{X}$ given diffusion constant $D$, is
\begin{equation} \label{Eq:Act}
\P{\mathbf{X}}{D} = \prod^{N}_{i=1} \P{\mathbf{x}_{i+1}}{\mathbf{x}_i}.
\end{equation}
In Eq.~\ref{Eq:Act}, $\P{\mathbf{x}_{i+1}}{\mathbf{x}_i}=\P{\mathbf{x}_{i+1}}{\mathbf{x}_i,D}$ is the probability density of each discrete jump from $\mathbf{x}_i\to\mathbf{x}_{i+1}$ over time step $\delta t_i=t_{i+1}-t_i$, given diffusion constant $D$.

When measured experimentally, however, the true positions $\mathbf{X}$ are never known exactly, but are related to $N$ observed positions $\mathbf{O}$ by some distribution $\P{\mathbf{o}_i}{\mathbf{x}_i,{\mathbf{x}_{i+1} }}$, where the dependence on both $\mathbf{x}_i$ and $\mathbf{x}_{i+1}$ arises from the effects of exposure time integration by the observation apparatus which will be dealt with in detail later.  Under this experimental model, $\P{\mathbf{O},\mathbf{X}}{D}$, the combined likelihood of the observed positions $\mathbf{O}$ and the actual positions $\mathbf{X}$ is a product of the observation probability densities  $\P{\mathbf{o}_i}{\mathbf{x}_i,\mathbf{x}_{i+1}}$ and the diffusion transition probability densities $\P{\mathbf{x}_{i+1}}{\mathbf{x}_i}$ for each of the $N$ observed positions and displacements,

\begin{equation}
\label{eq:poxd}
\P{\mathbf{O},\mathbf{X}}{D} =
\prod^{N}_{i=1} \P{\mathbf{o}_i}{\mathbf{x}_i,\mathbf{x}_{i+1}} \P{\mathbf{x}_{i+1}}{\mathbf{x}_i}.
\end{equation}
Since $\mathbf{X}$ is unknown for experimental data, we integrate Eq.~\ref{eq:poxd} over all possible $\mathbf{X}$ to marginalize out the dependence on $\mathbf{X}$,
and write the diffusion likelihood as an integral over the space of all $\mathbf{X}$-values,
\begin{equation*} \label{ndintegral2}
\P{\mathbf{O}}{D}
=\int \d{\mathbf{X}} \P{\mathbf{O},\mathbf{X}}{D}
=\int \d{\mathbf{X}} \prod^{N}_{i=1} \P{\mathbf{o}_i}{\mathbf{x}_i,\mathbf{x}_{i+1}} \P{\mathbf{x}_{i+1}}{\mathbf{x}_i}.
\end{equation*}

Experimental data typically involves trajectories with two or three spatial dimensions.  For diffusion in an isotropic medium and particle uncertainties given as normal distributions with no covariance among the spatial dimensions, the probability distribution of a particular displacement in each dimension is separable.  Thus, if $\Upsilon$ is the number of dimensions, then
\begin{equation}
\label{eq:separable}
\P{\mathbf{O}}{D} = \prod_{n=1}^\Upsilon \P{O_n}{D}.
\end{equation}
Hence, it is sufficient to only consider the estimation problem in the one-dimensional (1D) case $O=\{o_i\}_{i=1}^N$, and
\begin{equation} \label{eq:pod}
    \P{O}{D} = \intRNI \d{X} \prod^{N}_{i=1} \P{o_i}{x_i,x_{i+1}} \P{x_{i+1}}{x_i}.
\end{equation}

\subsection{Accounting for the effects of exposure time integration}
Equation~\ref{eq:pod} is the fundamental description of the likelihood of diffusion constant $D$ given observations $O$.  Unfortunately, solving for this expression explicitly is difficult because every $o_i$ term is dependent on both $x_i$ and $x_{i+1}$.  This is because the estimate of $o_i$'s position is typically made from data collected over an exposure time $0<t_\epsilon\leq t_{i+1}-t_i$.  If the observational apparatus is a camera sensor, the signal will be integrated over the frame, resulting in a motion-blurred image of the moving particle, hence the observed location is conditional upon the particle's true position at the beginning ($x_i$) and end ($x_{i+1}$) of the frame.

In the case where exposure time $t_\epsilon$ goes to 0, but $\delta t_i=t_{i+1}-t_i$ remains constant, the
motion-blur effect is no longer present, so the observed location $o_i$ depends only on position $x_i$,
\begin{equation}
    \label{eq:pod-pulsed}
    \P{O}{D} = \intRN \d{X} \prod^{N}_{i=1} \P{o_i}{x_i} \prod^{N-1}_{j=1} \P{x_{j+1}}{x_j}.
\end{equation}
Without the additional dependence on $x_{i+1}$, the methods required to solve the integral in Eq.~\ref{eq:pod-pulsed} are simpler.
In order to use this simpler representation, we will transform Eq.~\ref{eq:pod} into a form which resembles Eq.~\ref{eq:pod-pulsed}, and seek functions
$\M(o_i,x_i)$ and $\T(x_{j+1},x_j)$ such that
\begin{equation} \label{eq:PtoM}
    \P{O}{D} = \intRNI \d{X} \prod^{N}_{i=1} \P{o_i}{x_i,x_{i+1}} \P{x_{i+1}}{x_i} = \intRN \d{X} \prod^{N}_{i=1} \M(o_i,x_i) \prod^{N-1}_{j=1} \T(x_{j+1},x_j).
\end{equation}
The function $\T(x_{i+1},x_i)$ stands for the transition probability; it is simply the probability of a particle diffusing with constant $D$ moving from $x_i$ to $x_{i+1}$, over time $\delta t_i$.  The function $\M(o_i,x_i)$ stands for the measurement probability and it encapsulates the net effect of both the measurement localization error and the motion-blur.  The details of the representation equivalence of Eq.~\ref{eq:PtoM} are important for correctness, but they also unnecessarily complicate the exposition, and so can be found in Sec.~\ref{sec:ExpDeriv}.  Other authors~\cite{Savin2005,Berglund2010} have investigated the motion-blur effects of exposure time integration, and found that the effect can be approximated by an effective decrease in variance of the measurement localization error, dependent on diffusion constant $D$ and exposure time $t_\epsilon$.  Our derivations in Sec.~\ref{sec:ExpDeriv} agrees with the effective correction factor in \cite{Savin2005,Berglund2010}, and more importantly provides a form for the diffusion likelihood that is directly amenable to the solution techniques we employ in
Secs.~\ref{sec:theory-recursive},\ref{sec:theory-laplace},~and~\ref{sec:theory-markov}.

The result of the transformation of Eq.~\ref{eq:PtoM} is that the effective measurement function $\M_i$  and the transition function $T_i$ take the form of normalized Gaussians.  We use the notation
\begin{equation*}
\N(a,a_0,\eta)=\frac{1}{\sqrt{2\pi\eta}}\exp{\left[ -\frac{(a-a_0)^2}{2\eta} \right]}.
\end{equation*}
to represent the normalized Gaussian function with variance $\eta=\sigma^2$ centered around
mean $a_0$ considered as a function of $a,a_0,$ and $\eta$.  Using this notation, we can succinctly represent the measurement and transition functions as,
\begin{align}
    \label{eq:varT} \T_i = \T_i(x_{i+1},x_i) = & \N(x_{i+1},x_i,\omega_i(D)), \quad\textrm{for } 1\leq i \leq N-1, \textrm{ and }\\
    \label{eq:varM} \M_i = \M_i(o_i,x_i) = & \N(o_i,x_i,\eps_i(D)),\quad\textrm{for } 1\leq i\leq N.
\end{align}
The transition functions (Eq.~\ref{eq:varT}), are unaffected by the motion blur transformation and their Gaussian representation follows directly from the normally distributed displacements of diffusive processes, hence the variance is
\begin{equation*}
\label{eq:var-diffusion}
\omega_i(D)=2D\delta t_i.
\end{equation*}
For the measurement functions (Eq.~\ref{eq:varM}), the variance $\eps_i(D)$, is the variance due to measurement error, $v_i$, combined with a correction for the effect of motion-blur that is dependent on diffusion constant $D$ and exposure time $t_\epsilon$,
\begin{equation*}
\label{eq:var-measurement}
\eps_i(D)=v_i- D t_{\eps}/3.
\end{equation*}
where the factor of $1/3$ comes from the continuous limit integration of photon emissions for averaged Brownian trajectories (Sec.~\ref{sec:TimeAv}). It is important to note that the independence of $t_{\eps}$ and $\delta t_i$ allows for gaps in the trajectories, where $\delta t_i$ could span a duration of multiple frames but $t_{\eps}$ is the exposure time of a single frame.

The result is that Eq.~\ref{eq:PtoM} allows us to express the likelihood function exactly in a simple form that deals directly with variable localization error, motion-blur effects, and missing or irregularly spaced trajectory localizations,
\begin{equation} \label{eq:Funk}
    \P{O}{D} = \intRN \d{X} \prod^{N}_{i=1} \M_i \prod^{N-1}_{j=1} \T_j.
\end{equation}

\section{Recursive Method}
\label{sec:theory-recursive}
The notation for the transition and measurement functions allows us to define the likelihood function $\L(D)$, by writing Eq.~\ref{eq:Funk} in a form that emphasizes the dependencies on each marginalized position $x_i$,
\begin{equation} \label{eq:recursiveform}
\L(D)=\P{O}{D} =  \int\d{x_N} \M_N \int \d{x_{N-1}} \M_{N-1} \T_{N-1} \ldots
\int\d{x_2} \M_2 \T_2 \int\d{x_1} \M_1 \T_1.
\end{equation}

The form of Eq.~\ref{eq:recursiveform} leads to a direct recursive solution, taking into account the properties of integrals over products of normalized Gaussian functions. Define $\L_i$ as the sub-integrand of $\L(D)$ considering only the first $i$ observations,
\begin{equation}\label{eq:LRecurseDef}
\begin{aligned}
    \L_1(D,x_2) & = \intII \M_1 \T_1 \d{x_1}, \\
    \L_i(D,x_{i+1}) & = \intII \M_i \T_i \mathcal{L}_{i-1} \d{x_i}, \quad 2\leq i\leq N-1 \\
    \L(D) =\L_N(D) & = \intII \M_N \mathcal{L}_{N-1} \d{x_N}. \nonumber
\end{aligned}
\end{equation}
Now, consider that the integral of a product of two normalized Gaussians sharing a parameter $x$, with means and variances denoted by $c_i$ and $\varphi_i$ respectively, is itself a normalized Gaussian (Sec.~\ref{sec:NormalIdent}),
\begin{equation} \label{eq:convolve2}
    \intII \d{x} \prod^2_{i=1} \N(x,c_i,\varphi_i) = \N(c_1,c_2,\varphi_1+\varphi_2).
\end{equation}
Hence, $\L_1$ is a normalized Gaussian in parameter $x_2$,
\[ \L_1(D,x_2) = \int\d{x_1} \M_1 \T_1 = \N(o_1 , x_2, \eps_1 + \omega_1). \]
This implies that $\L_2(D,x_3)$ which is an integral over positions $x_2$, can now be written as an integral over three normalized Gaussians, all of which
share parameter $x_2$,
\begin{equation} \label{eq:L2recursive}
\L_2(D,x_3)=\int\d{x_2}\M_2\T_2 \int\d{x_1} \M_1 \T_1= \int\d{x_2} \M_2\T_2\L_1.
\end{equation}
Similarly, the integral of a product of three normalized Gaussians sharing the integrated parameter is itself a product of two normalized Gaussians (Sec.~\ref{sec:NormalIdent}),
\begin{equation} \label{eq:convolve3}
    \intII \d{x} \prod^3_{i=1} \N(c_i,x,\varphi_i) = \N(c_1,c_2,\varphi_1+\varphi_2)\N(c_3,c',\gamma),
\end{equation}
where,
\begin{align} \label{gensubs}
    c' = \frac{c_1 \varphi_2 + c_2 \varphi_1}{\varphi_1+\varphi_2}, \quad \text{and} \quad
    \gamma = \frac{\varphi_1\varphi_2 + \varphi_1\varphi_3 + \varphi_2\varphi_3}{\varphi_1 + \varphi_2}. \nonumber
\end{align}
Hence, applying Eq.~\ref{eq:convolve3} to Eq.~\ref{eq:L2recursive}, we find that
\[\L_2=\int\d{x_2} \\M_2T_2\L_1 =\N(o_1,o_2,(\eps_1+\omega_1)+\eps_2)\N(x_3,\mu_2,\eta_2) \]
is a product of two normalized Gaussians, one of which depends on $x_3$ and the other is a constant with respect to $X$.  The variables $\mu_2$ and $\eta_2$ follow from Eq.~\ref{eq:convolve2} and Eq.~\ref{eq:convolve3}, and are the second pair of values in a recursive solution.  Since all subsequent integrals, barring the last integral, can be expressed as a product of three normalized Gaussians, we can express the recursion variables as
\begin{equation}
\label{eq:rec-vars1}
    \mu_1 = o_1, \quad\ \eta_1 = \eps_1 + \omega_1, \quad\ \text{and} \quad \alpha_1 = \eta_1 + \eps_2,
\end{equation}
and for $2\leq i \leq N-1$,
\begin{equation}
\label{eq:rec-vars2}
    \mu_i = \frac{\mu_{i-1} \eps_i + \eta_{i-1} o_i  } {\alpha_{i-1} }, \quad \eta_i = \frac{ \eta_{i-1} \eps_i} {\alpha_{i-1}} + \omega_i, \quad \text{and} \quad \alpha_i = \eta_i+\eps_{i+1}.
\end{equation}
Finally, this allows us to express our integrands $\L_i$ as
\begin{equation}
\label{eq:recursive-sol}
\begin{aligned}
    \L_1 & = \N(x_2,\mu_1, \eta_1)  \\
    \L_i & = \N(x_{i+1},\mu_i,\eta_i) \prod_{k=1}^{i-1} \N(o_{k+1}, \mu_k, \alpha_k) , \quad 2\leq i\leq N-1 \\
    \L(D) = \L_N & = \prod_{k=1}^{N-1} \N(o_{k+1}, \mu_k, \alpha_k) .
\end{aligned}
\end{equation}
Equation~\ref{eq:recursive-sol} is the final form of the recursive solution for $\L(D)$ which is simply the product of $N-1$ normalized Gaussians each of which has parameters which come from a recursive relationship on $o_i$, $\eps_i$, and $\omega_i$.  The value of $D$ that maximizes $\L(D)$ is the maximum likelihood estimate.

\section{Laplace Method}
\label{sec:theory-laplace}
An independent solution for Eq.~\ref{eq:Funk} can be obtained using the Laplace method which is based on integrating the second moment of the Taylor expansion of the exponential component of a function.  Given that the second moment of a Taylor expansion is quadratic, this ensures that the function under the integral is always a Gaussian function \cite{de1774memoire}. Another caveat to the Laplace method is that the Taylor expansion has to occur about the peak of the exponential, so that the first moment of the Taylor expansion goes to 0.  To perform the Laplace method, we express our likelihood $\L(D)=\P{O}{D}$ in terms of exponential and non-exponential components
\begin{equation*}
    \label{eq:laplaceform}
    \L(D) = \intII \d{X} f(X) = \intII \d{X} h(X) \Exp{-g(X)},
\end{equation*}
where $f(x)$ is simply the integrand of Eq.~\ref{eq:Funk},
\begin{equation}
    \label{eq:LaplaceF}
    f(X)=h(X) \Exp{-g(X)}=\prod_{i=1}^{N}\M_{i}\prod_{j=1}^{N-1}\T_{j}.
\end{equation}
Thus, using equations~\ref{eq:varM}~and~\ref{eq:varT}, we see that $h=h(X)$ is independent of $X$ and $g(X)$ is quadratic in $X$,
\begin{align*}
    h = & \prod^N_{i=1}  \frac{1}{\sqrt{2 \pi \eps_i}}  \prod^{N-1}_{i=1}  \frac{1}{\sqrt{2 \pi  \omega_i}},\\
    g(X) = & \sum^N_{i=1} \frac{(o_i-x_i)^2}{2 \eps_i} + \sum^{N-1}_{i=1} \frac{(x_{i+1}-x_i)^2}{2 \omega_i}.
\end{align*}

The maximum likelihood estimate $\Xhat$ of the actual positions $X$, given $D$ and $O$ will be wherever the integrand is maximized, and since $g(X)\geq0$,
\begin{equation*}
\Xhat = \argmax_X f(X) = \argmin_X g(X).
\end{equation*}

Now, given that $g(X)$ is quadratic, a second order Taylor expansion of $g$ about $\Xhat$ is exact and the Laplace method will provide an exact solution
for $\L(D)$ as the integral can be shown to take the form of a standard Gaussian integral.  To see this, we write out the second order Taylor expansion
\begin{equation}
    \label{eq:Dlaplaceform}
    \intII \d{X} f(X) = h \intII \d{X} \Exp{-g(\Xhat) - \nabla g(\Xhat)(X-\Xhat) - \frac{1}{2} (X-\Xhat)^\transp \nabla \nabla g(\Xhat) (X-\Xhat) }.
\end{equation}
Since $\Xhat$ is the minima of $g(X)$, the gradient $\nabla g(\Xhat)=0$, and we can rearrange
Eq.~\ref{eq:Dlaplaceform} to extract all terms independent of $X$,
\begin{equation} \label{eq:laplaceint}
    \intII \d{X} f(X) = f(\Xhat) \intII \d{X} \Exp{- \frac{1}{2} (X-\Xhat)^\transp \nabla \nabla g(\Xhat) (X-\Xhat) }.
\end{equation}
Furthermore, since $h$ is independent of $X$, we know that $-\nabla\nabla \ln{f(X)} = \nabla \nabla g(X)=M$, where $M$ can be thought of as the inverse of the covariance matrix for the multivariate Gaussian, or equivalently as the Hessian matrix of $-\ln{f(X)}$.  Substituting $M$ for $\nabla \nabla g(X)$ in Eq.~\ref{eq:laplaceint} we are left with a Gaussian integral with the solution,
\begin{equation}
    \label{eq:laplacelikelihood}
    \L(D) = f(\Xhat) \intII \d{X} \Exp{- \frac{1}{2} (X-\Xhat)^\transp M (X-\Xhat) }  = f(\Xhat) \sqrt{ \frac{(2\pi)^N}{\det{M}}}.
\end{equation}
The Hessian matrix $M$ is independent of $X$ and is symmetric tri-diagonal with non-zero elements,
\begin{equation}
\label{eq:LaplaceHessian}
\begin{aligned}
    M_{1,1}   &= -\frac{\partial^2\ln f}{\partial x_1^2} = \frac{1}{\eps_1}+\frac{1}{\omega_1}  \\
    M_{i,i}   &= -\frac{\partial^2\ln f}{\partial x_i^2} = \frac{1}{\eps_i}+\frac{1}{\omega_i}+\frac{1}{\omega_{i-1}}, \quad 2\leq i \leq N-1  \\
    M_{N,N}   &= -\frac{\partial^2\ln f}{\partial x_N^2} = \frac{1}{\eps_N}+\frac{1}{\omega_{N-1}}  \\
    M_{i,i+1} = M_{i+1,i} &= -\frac{\partial^2\ln f}{\partial x_i\partial x_{i+1}} = -\frac{1}{\omega_i}, \quad 1\leq i \leq N-1.
\end{aligned}
\end{equation}
We also require $\Xhat$ to compute Eq.~\ref{eq:laplacelikelihood}, which can be solved for with the relation $-\nabla \ln f(\Xhat)=0$ (Sec.~\ref{sec:MethCom}),
giving
\begin{equation} \label{eq:MLE_X}
    \Xhat = M^{-1} \Theta,
\end{equation}
where $\Theta=\{\theta_i = o_i/\eps_i\}_{i=1}^N$.

\section{Markov Method}
\label{sec:theory-markov}
In this section we present a derivation of the likelihood function $\L(D)$ utilizing a technique developed by Andrey Markov \cite{markov1912} and generalized by Chandrasekhar \cite{Chandrasekhar1943}.  Markov's method allows us to transform $\P{O}{D}$ (Eq.~\ref{eq:Funk}) from a function of the $N$ observed positions $O=\{o_i\}_{i=1}^N$ into a function of the $N-1$ discrete steps (displacements) between subsequent observations
\begin{equation*}
\Jump=\{\jump_i = o_{i+1}-o_i\}_{i=1}^{N-1}.
\end{equation*}
This is possible because the spatial invariance of diffusion means $\L(D)$ depends not on the absolute spatial positions $O$, but only their relative displacements, $\Jump$.  Thus we should expect that $\P{O}{D}$ can also be expressed as $\P{\Jump}{D}$, however Eq.~\ref{eq:recursiveform}, which defines $\P{O}{D}$, cannot be directly transformed into a function on $\Jump$.  This is where Markov's method allows us to solve for a function $\P{\Jump}{D}=\P{O}{D}$ for a given $D$ value.  For a particular fixed $\Jump$ and $\d{\Jump}$ of interest, the value of $\P{\Jump}{D}\d{\Jump}$ gives the probability that variable $\Jump'$ is within the bounds
\begin{equation}
\label{eq:dS-bounds}
\Jump-\frac{1}{2}\d{\Jump}\leq \Jump' \leq \Jump+\frac{1}{2}\d{\Jump}.
\end{equation}
More formally, $\P{\Jump}{D}\d{\Jump}$ is the integral over the volume $\d{\Jump}$ around the point of interest $\Jump$, and we let $\Jump'$ represent
the variable integrated over,
\begin{equation*}
\P{\Jump}{D}\d{\Jump} = \int^{\Jump + \frac{1}{2}\d{\Jump}}_{\Jump - \frac{1}{2}\d{\Jump}} \d{\Jump'} \P{O}{D}.
\end{equation*}
The issue remains that $\P{O}{D}$ is expressed in a basis of $O$ rather than of $S$, and integrating with respect to bounds in a different basis is non-trivial.  In order to circumvent this issue, Markov utilized a product of Dirichlet integrals, $\Xi(\Jump') = \prod^{N-1}_{k=1} \xi_k(\jump_k')$, to expand the limits of integration to all space

\begin{equation} \label{Markov1}
    \P{\Jump}{D} \d{\Jump}= \intII \d{\Jump'} \Xi(\Jump') \P{O}{D}.
\end{equation}
The idea is that for each dimension of $S$, the Dirichlet integral $\xi_k(\jump_k')$ acts like a continuous indicator function determining if $\jump_k'$ is within the bounds of Eq.~\ref{eq:dS-bounds},
\begin{equation*} \label{dirichletInt}
    \xi_k(\jump_k') = \frac{1}{\pi} \intII  \d{\rho_k} \frac{\sin(\frac{1}{2}\d{\jump_k} \rho_k)}{\rho_k} \Exp{\imath \rho_k (\jump_k' - \jump_k) }, \nonumber
\end{equation*}
so that,
\begin{equation}
    \xi_k(\jump_k') =
    \begin{cases}
        1 &  \jump_k-\frac{1}{2}\d{\jump_k}\leq \jump_k' \leq \jump_k+\frac{1}{2}\d{\jump_k} \\
        0 & \text{otherwise}
    \end{cases}.\nonumber
\end{equation}
Therefore $\Xi(\Jump')$ is the indicator function acting over the whole space of $S'$, and determining if $S'$ is within the volume $\d{S}$ around our point of interest $S$,
\begin{equation}
    \Xi(\Jump') = \prod_{k=1}^{N-1} \xi_k(\jump_k') =
    \begin{cases}
        1 &  \displaystyle \bigwedge_{k=1}^{N-1} \textstyle s'_k \in \left[\jump_k-\frac{1}{2}\d{\jump_k}, \jump_k+\frac{1}{2}\d{\jump_k}\right] \\
        0 & \text{otherwise}
    \end{cases}.\nonumber
\end{equation}

This puts Eq.~\ref{Markov1} in the form
\begin{equation}
    \label{eq:expanded-rho-S-int}
    \P{\Jump}{D} \d{\Jump}= \int \d{\Jump'} \frac{1}{\pi^{N-1}} \int \d{\rhovec} \left[ \prod^{N-1}_{k=1} \frac{\sin(\frac{1}{2} \d{\jump_k} \rho_k)}{\rho_k} \right] \Exp { \imath \rhovec^\transp (\Jump' - \Jump) } \P{O}{D},
\end{equation}
where $\rhovec =\{\rho_k\}_{k=1}^{N-1}$ is a vector of conjugate coordinates to $\Jump'$.  We can then rearrange Eq.~\ref{eq:expanded-rho-S-int}
to move the integral over $\d{\Jump'}$ and all factors dependent on $\Jump'$ into function $\Lambda({\rhovec})$,
\begin{equation} \label{eq:MarkovTrans}
    \P{\Jump}{D} \d{\Jump}= \frac{1}{\pi^{N-1}} \int \d{\rhovec} \left[ \prod^{N-1}_{k=1} \frac{\sin(\frac{1}{2} \d{\jump_k} \rho_k)}{\rho_k} \right]
    \Exp {-\imath \rhovec^\transp \Jump } \Lambda(\rhovec).
\end{equation}
Now, we can interpret $\Lambda({\rhovec})$ as the characteristic function of $\P{O}{D}$ in the $\Jump'$ basis and it has the form
\begin{equation} \label{fourierTerm}
    \Lambda(\rhovec) = \int \d{\Jump'} \Exp{\imath\rhovec^\transp \Jump' } \P{O}{D}.
\end{equation}
The form of Eq.~\ref{fourierTerm} implies that $\Lambda(\rhovec)$ is the inverse Fourier transform of $\P{O}{D}$.
Due to the properties of the Fourier transform, we assert $\Lambda(\rhovec)$ is a bounded function with a finite integral because $\P{O}{D}$ is expressed as a finite product of Gaussians with non-zero variance, hence it is a continuous function \cite{cahill2013physical}.
Given that $\int \d{\rhovec} \Lambda(\rhovec)$ is bounded and $\d{\Jump}$ is small we can approximate the product of sinc functions in Eq.~\ref{eq:MarkovTrans} as
\begin{equation*}
\prod_{k=1}^{N-1} \frac{\sin(\frac{1}{2}\d{\jump_k} \rho_k)}{\rho_k}=\prod_{k=1}^{N-1} \frac{\d{\jump_k}}{2}=\frac{\d{\Jump}}{2^{N-1}}.
\end{equation*}
Thus Eq.~\ref{eq:MarkovTrans} becomes the Fourier transform of $\Lambda(\rhovec)$
\begin{equation} \label{MarkovForm}
    \P{\Jump}{D} \d{\Jump}= \frac{\d{\Jump}}{(2\pi)^{N-1}}\int \d{\rhovec} \Exp{-\imath\rhovec^\transp\Jump}\Lambda(\rhovec).
\end{equation}

We are now interested in evaluating $\Lambda(\rhovec)$ explicitly. To do so we note that $\int \d{\Jump'} \P{O}{D}=1$ since $\P{O}{D}$ is a probability
distribution that, as we have argued, can be equivalently expressed in the $S$ basis as $\P{S}{D}$ (Sec.~\ref{sec:MethCom}).
With this understanding we evaluate $\Lambda(\rhovec)$, by expanding the exponential under integration in Eq.~\ref{fourierTerm} as a
Taylor series about the origin
\begin{equation}
    \Exp{\imath \rhovec^\transp \Jump'} = 1 + \imath \left( \sum^{N-1}_{j=1} \rho_j \jump_j' \right)
     -\frac{1}{2}\left( \left[\sum^{N-1}_{j=1} \rho_j^2 \jump_j'^{\,2} \right] + \left[ \sum^{N-2}_{j=1} \sum^{N-1}_{k=j+1} 2\rho_j\rho_k \jump'_j\jump'_k \right] \right) + \mathcal{O}(\rhovec^3). \nonumber
\end{equation}

Solving the integral for $\Lambda(\rhovec)$ with the Taylor expanded exponential term given that we know $\P{O}{D}$ is a normalized probability density under $\Jump'$ allows us to write $\Lambda(\rhovec)$ in terms of expected values for $s_k$ (using $\E{\cdot}$ to represent expectation),
\begin{equation} \label{sumofExpect}
    \Lambda(\rhovec)  =  1 + \imath \left( \sum^{N-1}_{j=1} \rho_j \E{\jump_j'} \right)
     -\frac{1}{2}\left( \left[\sum^{N-1}_{j=1} \rho_j^2 \E{\jump_j'^{\,2}} \right] + \left[ \sum^{N-2}_{j=1} \sum^{N-1}_{k=j+1} 2\rho_j\rho_k \E{\jump'_j\jump'_k} \right] \right)
     + \mathcal{O}(\rhovec^3) .
\end{equation}

We take the approach of solving for the expectation values on $\Jump'$ to see if those results will simplify the expression in Eq.~\ref{sumofExpect}.  If we set one of the observations, $o_i$, as a constant, we find that we can marginalize all of $\P{O}{D}$ except for the terms that are independently represented by the basis we are interested in.  In other words, we find that
\begin{align}
    \E{\jump_i'} & = \int \d{\jump_i'}  \jump_i' \N(\jump_i',0,\eps_i + \eps_{i+1} + \omega_i) = 0 \nonumber \\
    \E{\jump_i'^{\,2}} & = \int \d{\jump_i'}  \jump_i'^{\,2} \N(\jump_i',0,\eps_i + \eps_{i+1} + \omega_i) = \eps_i + \eps_{i+1} + \omega_i. \nonumber \\
    \E{\jump_i' \jump_{i+1}'} & = \int \d{\jump_i'} \d{\jump_{i+1}'} \jump_i' \jump_{i+1}' \N(S_b,0,\Sigma_b) = -\eps_{i+1}. \nonumber
\end{align}
Where the substitution of variables required for integrating the expectation of $\E{\jump_i' \jump_{i+1}'}$ induces a bivariate Gaussian function with location parameters $S_b = [s_i, s_{i+1}]^\transp$ and covariance matrix
\begin{equation}
    \Sigma_b = \begin{bmatrix} \omega_i + \epsilon_i + \epsilon_{i+1} & -\epsilon_{i+1} \\
        -\epsilon_{i+1} & \omega_{i+1} + \epsilon_{i+1} + \epsilon_{i+2} \end{bmatrix}. \nonumber
\end{equation}

Furthermore, we find that the separability of two non-adjacent displacements results in the relation $\E{\jump_i' \jump_{i+k}'} = \E{\jump_i'}\E{\jump_{i+k}'} = 0 $ for $k>1$.  Since $\P{O}{D}$ is a multivariate Gaussian with 0 mean, we can apply Isserlis's theorem to solve for all of the moments of the distribution \cite{Isserlis1918} given knowledge of the second moments of the Fourier transform on $\P{O}{D}$, which can be expressed as a covariance matrix.  This allows us to express Eq.~\ref{sumofExpect} as
\begin{align} \label{fourierCompact}
    \Lambda(\rhovec) = \Exp {- \frac{1}{2} \rhovec^\transp \Sigma \rhovec }.
\end{align}
The covariance matrix $\Sigma$ is symmetric tri-diagonal, with non-zero elements
\begin{equation}
\label{eq:markov-cov-mat}
\begin{aligned}
    & \Sigma_{i,i} = \omega_i + \eps_i + \eps_{i+1} \\
    & \Sigma_{i,i+1} = \Sigma_{i+1,i}= - \eps_{i+1}.
\end{aligned}
\end{equation}

The expression in Eq.~\ref{fourierCompact} is well known as the characteristic function of a multivariate Gaussian.  Substituting  Eq.~\ref{fourierCompact} into Eq.~\ref{MarkovForm} and factoring out $\d{\Jump}$ gives
\begin{align} \label{eq:markov-final}
    \L(D) = \P{\Jump}{D} = \frac{1}{\sqrt{(2\pi)^{N-1}\det{\Sigma}}} \Exp {-\frac{1}{2} \Jump^\transp \Sigma^{-1} \Jump }.
\end{align}

\section{Implementation}
\label{sec:implementation}
We have presented three independent solutions to the experimental diffusion likelihood $\L(D)$ (Eq.~\ref{eq:recursiveform}): the recursive method (Sec.~\ref{sec:theory-recursive}), the Laplace method (Sec.~\ref{sec:theory-laplace}), and the Markov method (Sec.~\ref{sec:theory-markov}).  While each method requires separate consideration, several features are common to all of the implementations.  The separability of the problem allows us to estimate diffusion constants for any dimensional inputs inputs using the 1D algorithms (Eq.~\ref{eq:separable}).  The inputs to the algorithms are: (1) the observed particle locations, $\mathbf{O}=\{\mathbf{o}_i\}_{i=1}^N$; (2) the observation times $T=\{t_i\}_{i=1}^N$;
(3) the measurement variance for each observation $\mathbf{V}=\{\mathbf{v}_i\}_{i=1}^N$; (4) the exposure of each frame $t_{\epsilon}$; and (5) one or more diffusion constants $D$ at which to evaluate the likelihood.  The output for each $D$ value is $\ln(\L(D))$.  The logarithm of the likelihood makes the computation of products and exponentials much faster, and avoids the problem of numerical underflow for very small values of $\L(D)$.  Additionally, because the logarithm is a strictly monotonically increasing function, $\argmax_D{\L(D)}=\argmax_D{\ln(\L(D))}$, so the maximum likelihood estimate is identical for the log-likelihood.

\subsection{Recursive Method}

The recursive algorithm follows directly from the recursively defined variables
(Eqs.~\ref{eq:rec-vars1}~and~\ref{eq:rec-vars2}), and the expression of $\L(D)$
as a product of Gaussians (Eq.~\ref{eq:recursive-sol}).  The recursive expressions for $\alpha_i$, $\eta_i$, and $\mu_i$, are causal (the $i$-terms depend only on the $(i-1)$-terms), enabling their computation in a simple for loop over $N$.  Noting that the logarithm of a normalized Gaussian is
\begin{equation}
\label{eq:lognormal}
\ln\N(a,b,v)=-\frac{1}{2}\left[\ln(2\pi)+\ln(v)+\frac{(a-b)^2}{v}\right],
\end{equation}
we apply Eq.~\ref{eq:lognormal} directly to Eq.~\ref{eq:recursive-sol} to arrive at a computationally efficient form for the recursive solution of the log-likelihood
\begin{equation}
\ln\L(D)=\sum_{i=1}^{N-1} \ln\N(o_{i+1}, \mu_i, \alpha_i) = -\frac{1}{2}\left[(N-1)\ln(2\pi)+\sum_{i=1}^{N-1} \ln(\alpha_i)+\sum_{i=1}^{N-1} \frac{(o_{i+1}-\mu_i)^2}{\alpha_i} \right].\nonumber
\end{equation}
Of all the methods, the recursive method is the simplest to implement and the most computationally efficient and numerically stable.

\subsection{Laplace Method}
The computational core of the Laplace method centers around the Hessian matrix $M$ (Eq.~\ref{eq:LaplaceHessian}).  This matrix is symmetric tri-diagonal, which means all non-zero elements are on the main diagonal and the diagonals immediately above and below.  Using $M$ we can solve the linear system $\Xhat = M^{-1} \Theta$ (Eq.~\ref{eq:MLE_X}) to obtain the maximum likelihood estimates $\Xhat$ for the true particle locations.  Typically, solving large linear systems is expensive but since $M$ is tri-diagonal there are algorithms to solve this system in linear time~\cite{el2004inverse}.  We refer the reader to Sec.~\ref{sec:Imp} for the details of tri-diagonal matrix algorithms and our implementation.

Given a solution for $\Xhat$, we can use the definition of $f(X)$ in Eq.~\ref{eq:LaplaceF}
along with Eq.~\ref{eq:lognormal} to compute
\begin{equation}
\ln f(\Xhat)=-\frac{1}{2}\left[
\sum_{i=1}^{N} \ln(2\pi\eps_i)
+ \sum_{i=1}^{N} \frac{(o_i-\widehat{x}_i)^2}{\eps_i}
+ \sum_{i=1}^{N-1} \ln(2\pi\omega_i)
+ \sum_{i=1}^{N-1} \frac{(\widehat{x}_{i+1}-\widehat{x}_i)^2}{\omega_i} \right].
\end{equation}
Finally we can compute the log-likelihood using the Laplace solution of Eq.~\ref{eq:laplacelikelihood}, finding that
\begin{equation}
\label{eq:LaplaceLLH}
\begin{aligned}
\ln\L(D)&= \ln f(\Xhat)+\frac{N}{2}\ln(2\pi)-\frac{1}{2}\logdet M\\
&= -\frac{1}{2}\left[(N-1)\ln(2\pi)
+ \sum_{i=1}^{N-1} \ln(\omega_i)
+ \sum_{i=1}^{N-1} \frac{(\widehat{x}_{i+1}-\widehat{x}_i)^2}{\omega_i}
+ \sum_{i=1}^{N} \ln(\eps_i)
+ \sum_{i=1}^{N} \frac{(o_i-\widehat{x}_i)^2}{\eps_i}
+\logdet M  \right]
\end{aligned}
\end{equation}

\subsection{Markov Method}
Finally, the Markov method computation, like the Laplace method, is centered around matrix computations.  In this case, the matrix of interest is the $N-1$ dimensional covariance
matrix $\Sigma$ (Eq.~\ref{eq:markov-cov-mat}), which also happens to be symmetric tri-diagonal, so the
same linear-time algorithms used in the Laplace method are applicable (Sec.~\ref{sec:Imp}).

For the Markov method computation we first solve the linear system $\Phi=\Sigma^{-1} S$,
then apply this solution along with the tri-diagonal log-determinant algorithm to compute the logarithm of the likelihood expression from
Eq.~\ref{eq:markov-final}, giving

\begin{equation}
\ln\L(D)=-\frac{1}{2}\left[(N-1)\ln(2\pi)
+ S^\transp \Phi
+ \logdet{\Sigma}
\right].\nonumber
\end{equation}

\section{Results}
\label{sec:results}

To demonstrate the benefits of including individual localization errors in the estimation, we opt for a loss function to evaluate the quality of the maximum likelihood estimators (MLE) under consideration. The mean squared error is a popular choice for evaluating estimators, but it is not a sufficient loss function for the estimation of $D$, where the mean squared error shows over penalization for higher estimations as it has a bounded loss at $\hat{D}=0$ and an unbounded loss at $\hat{D}=\infty$.  Instead, we will use the squared log loss function~\cite{brown1968inadmissibility}
\begin{equation*} \label{sqlogloss}
    \ell(D,\hat{D}) = \left( \ln(D) - \ln(\hat{D}) \right)^2.
\end{equation*}
The squared log loss function is similar to the mean squared error, except that the squared distance is between the logarithms of $D$ and $\hat{D}$.  From \cite{gelman2014bayesian}, we see that the variance term, $D$, scales with the observed data as a logarithm, hence the choice of the squared log loss function represents a metric between the expected data given by $D$ and $\hat{D}$.

For one set of SPT simulations, the true trajectory coordinates were first generated with the pure diffusion model.  Full frame motion blur was accounted for and localization errors were independently and identically drawn from either a Uniform or a Gamma distribution to test the effects of variable localization errors without attributing success to a particular choice of distribution.  As a control, a set of simulations with a constant localization error for all observations was generated to show that the likelihood distribution represented by the class of diffusion estimators that only recognize a Scalar Localization Error (SLE) was exactly the same as the distribution represented by either of our three derived methods which recognize a Vector of Localization Errors (VLE).  We then used the Gamma and Uniform distribution based localization errors to compare the new VLE based estimators that account for individually measured localization errors over the SLE based estimators, for which we had input the square root of the mean localization variance as the best representation for the scalar error in all trajectory observations.  We performed 10,000 trajectory trials for each data point to estimate the risk of using a particular estimator, where risk is defined as $R(D,\hat{D}) = \E{\ell(D,\hat{D})}$.  The squared log risk is the variance of $\ln(\hat{D})$ for an unbiased estimator.

The simulations with the gamma distributed localization errors were generated according to
\begin{equation*}
\label{eq:variablegV}
\sqrt{V}=\sigma\sim \mathrm{Gamma}(4,\langle{\sqrt{V}}\rangle/4),
\end{equation*}
where the standard gamma distribution p.d.f. is $\mathrm{Gamma}(k,\theta)= \theta^{-k} \sqrt{V}^{k-1} \exp(-\sqrt{V}/\theta)/\Gamma(k)$, and $\langle{\sqrt{V}}\rangle$ represents the mean error.  The simulations with the uniform distributed localization errors were generated according to
\begin{equation*}
\label{eq:variableuV}
\sqrt{V}=\sigma\sim \mathrm{Uniform}(\frac{1}{2} \langle{\sqrt{V}}\rangle, \frac{3}{2} \langle{\sqrt{V}}\rangle),
\end{equation*}
where the uniform distribution p.d.f. is $\mathrm{Uniform}(a,b) = 1/(b-a)$ for $\sqrt{V} \in [a,b]$ and $\mathrm{Uniform}(a,b) = 0$ for all other values of $\sqrt{V}$.
\begin{figure}[H]
\begin{center}
\begin{tabular}{ll}
(A) & (B)\\
\includegraphics[height=2.2in]{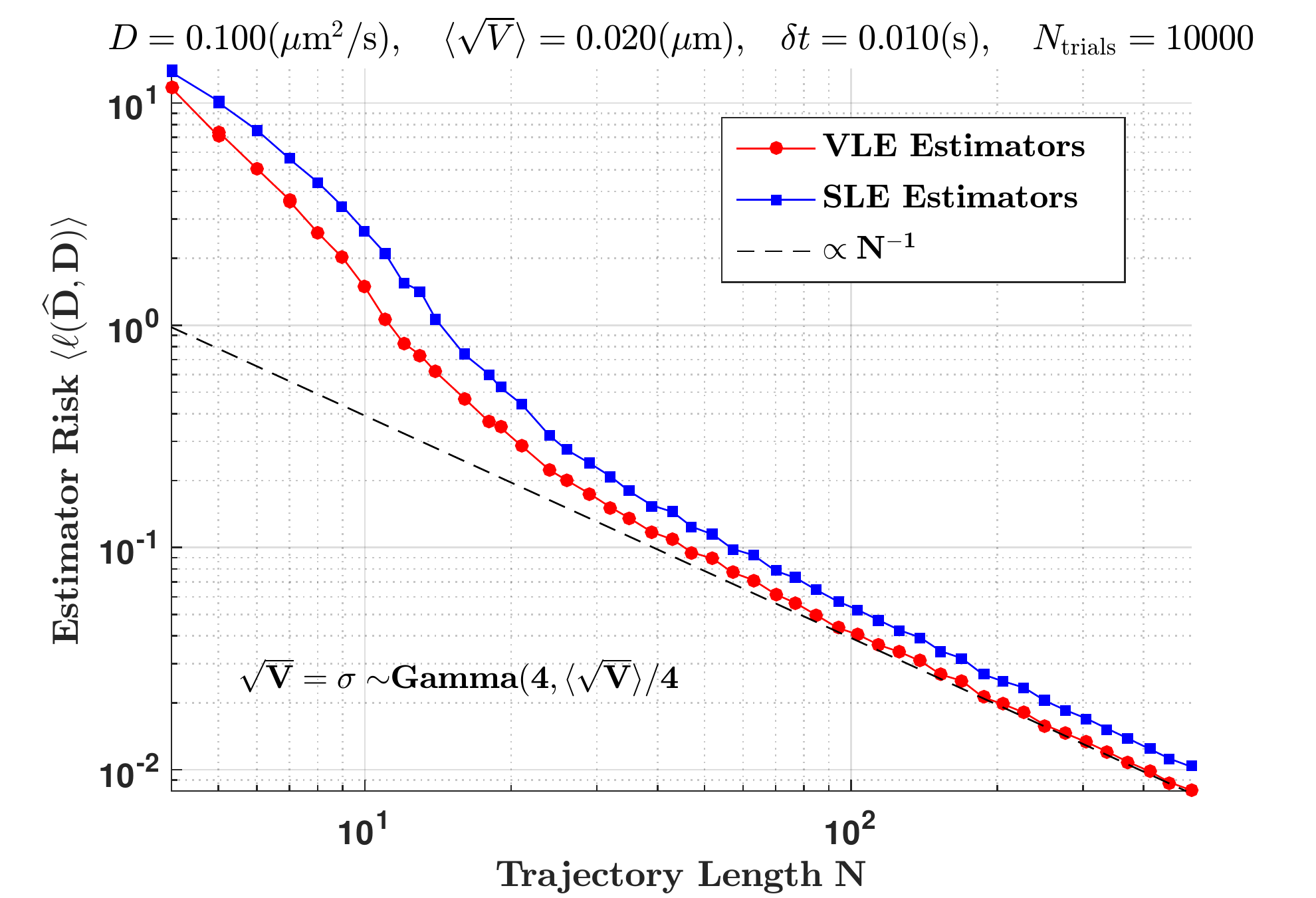}  & \includegraphics[height=2.2in]{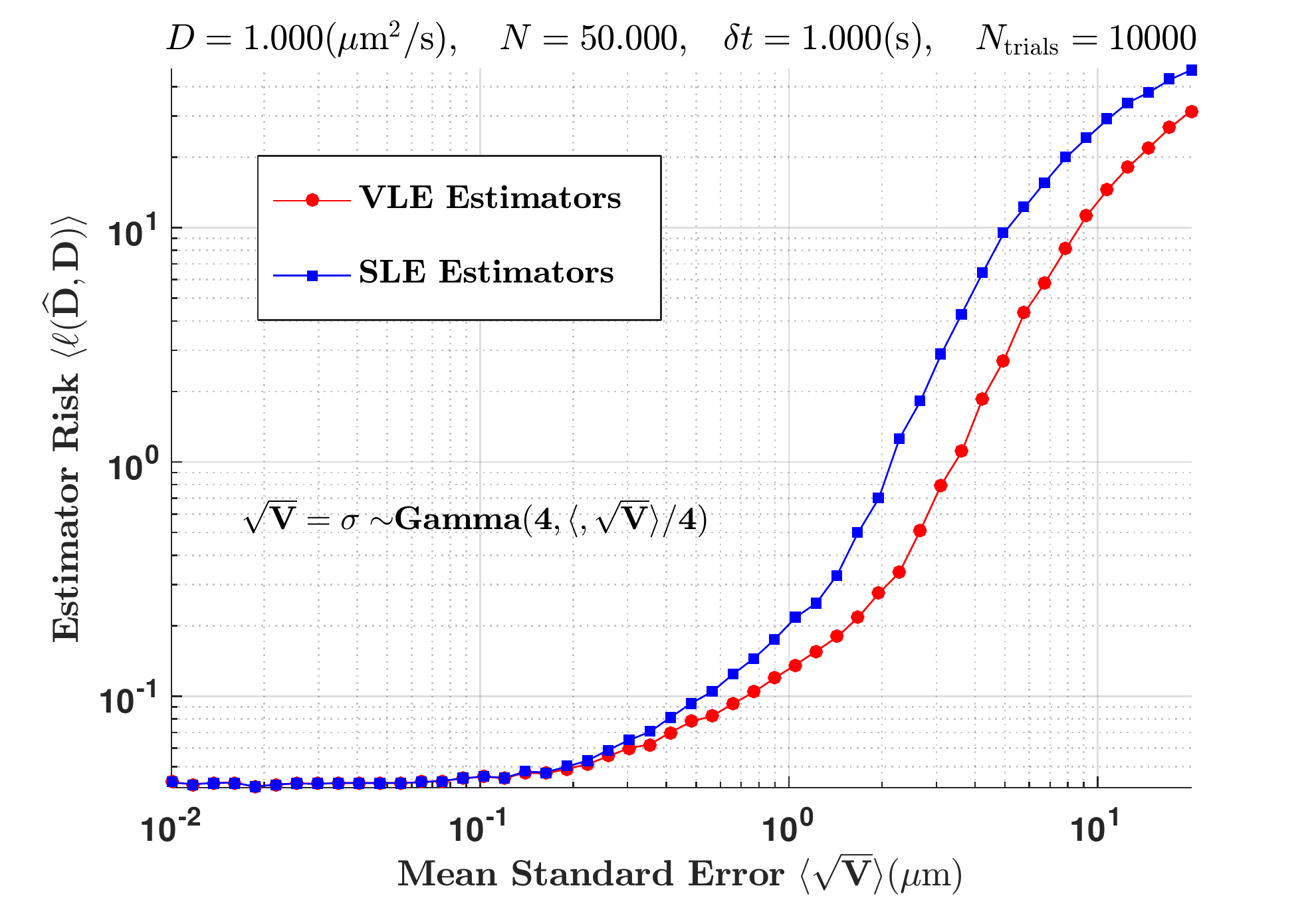} \\
(C) & (D)\\
\includegraphics[height=2.2in]{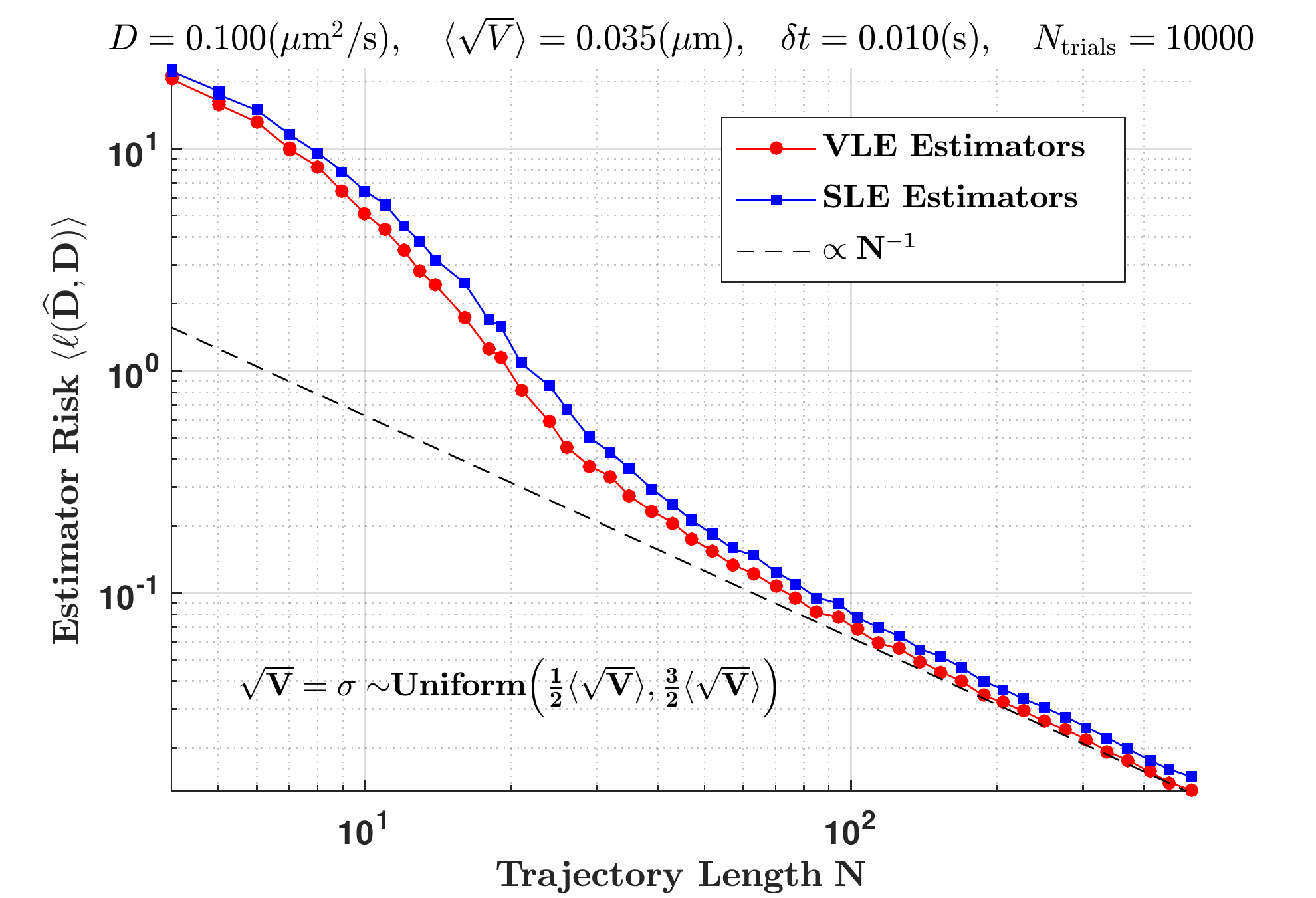}  & \includegraphics[height=2.2in]{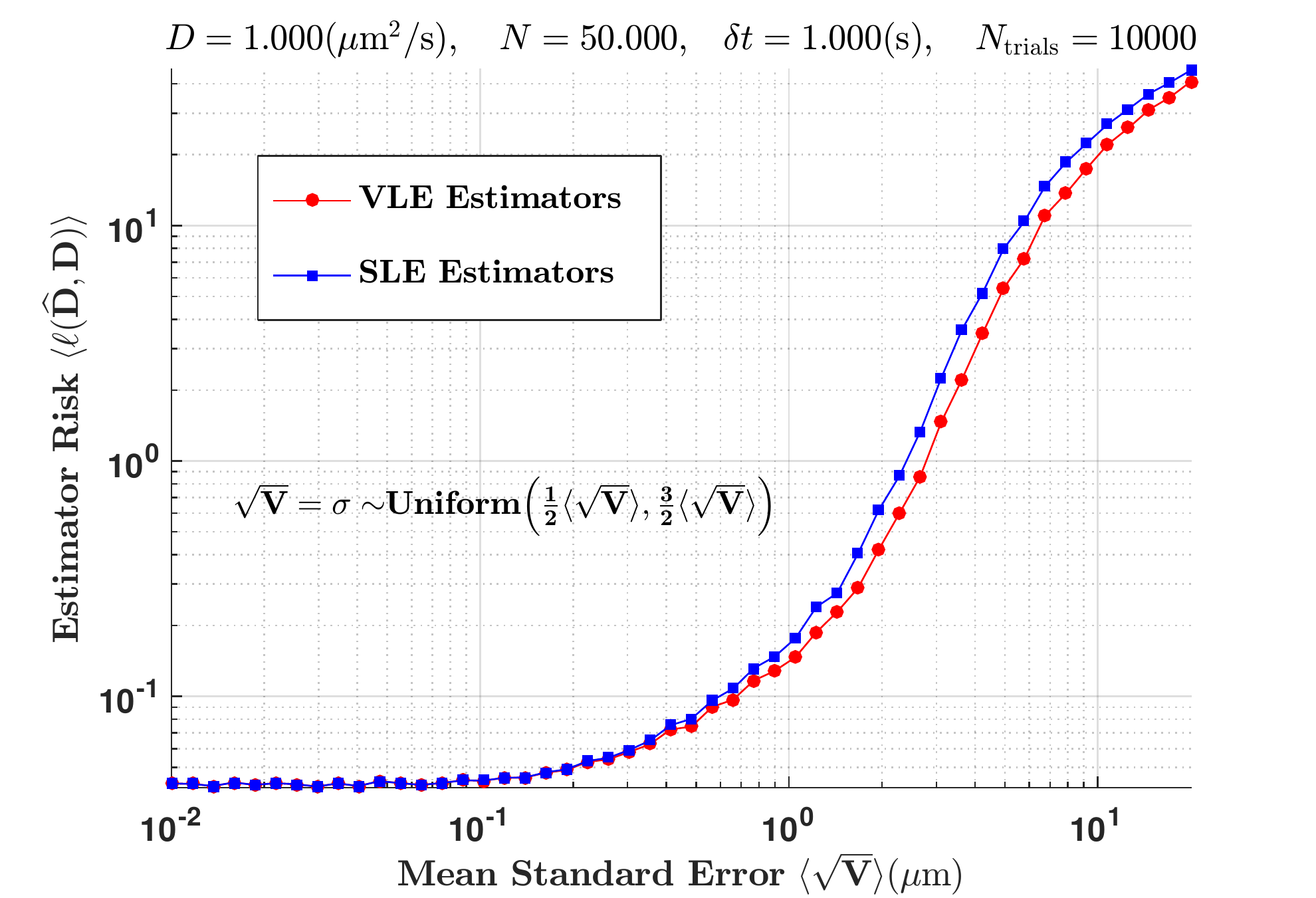}
\end{tabular}
\end{center}
\caption{Comparison of estimation risk of the Vector of Localization Errors (VLE) and Scalar Localization Error (SLE) MLEs with localization errors drawn from Gamma or Uniform distributions.  The MLE was found with the fminbnd command from Matlab on the calculated likelihood distributions with a lower bound of $10^{-8}$.  Log-log plots (A) and (C) show the risk of using a particular estimator on trajectories of various lengths with other constant parameters set to typical experimental values with standard errors drawn from gamma (A) or uniform (C) distributions. Log-log plots (B) and (D) show the risk of using a particular estimator on trajectories of length 50 with various standard errors drawn from gamma (B) or uniform distributions and all other parameters were set to 1 to study the effects of relative localization error.  In both (A) and (C) a fiducial line shows that after 30 observations the risk for the methods decreases approximately $\propto 1/N$ with trajectory length; in this regime the SLE estimators perform worse by a constant factor (to simulation precision) relative to the VLE estimators.  In (B) and (D) the risk for SLE estimators increase at a faster rate than the VLE estimators, indicating that the localization errors provide valuable information for improving estimator reliability.}
\label{fig:risks}
\end{figure}

Figures \ref{fig:risks}A and C are shown in log space to show that in the presence of sufficient information the risk of the VLE estimators is less than the risk of the SLE estimators by a constant proportionality factor when a scalar localization error is used to parameterize a continuous distribution of localization errors.  In other words, for a given parameterized scalar localization error, the amount of observations required to generate the same quality $\hat{D}$ estimate is always less by a proportional factor for the VLE estimators method compared to the SLE estimators.  Figures \ref{fig:risks}B and D are shown in log space to show how each estimator begins to fail in the presence of increasing relative localization error given a fixed set of trajectory observations (50).  For these subplots, the VLE estimators show a noticeable improvement in estimator reliability when the relative error is equal to or greater than the true underlying $D$.

In an experimental trajectory, the distribution that parameterizes the localization error is typically a function of several environmental variables so that it can often appear arbitrary or specific to a particular experimental trial.  In this manuscript, we focus on two simple distributions to provide a fair metric for validation over thousands of simulated trajectories.  It is worthwhile to further investigate the the precision increase of the VLE estimators over their SLE counterparts with localization error distributions of varying error variances.  We do so by performing trials on trajectories with localization errors parameterized by the Gamma and Uniform distribution, but this time we vary the parameters that characterize the variance of these distributions without altering the value of the mean localization variance.  To do so, we run simulations where the shape parameter, k, of the gamma distribution is altered so that our expression looks like
\begin{equation*}
\label{eq:variableKV}
\sqrt{V}=\sigma\sim \mathrm{Gamma}(k,\langle{\sqrt{V}}\rangle/k),
\end{equation*}
and the bounds of the uniform distribution are altered so the expression becomes
\begin{equation*}
\label{eq:variableuV}
\sqrt{V}=\sigma\sim \mathrm{Uniform}([1-b] \langle{\sqrt{V}}\rangle, [1+b] \langle{\sqrt{V}}\rangle).
\end{equation*}

\begin{figure}[H]
\begin{center}
\begin{tabular}{ll}
(A) & (B) \\
\includegraphics[height=2.2in]{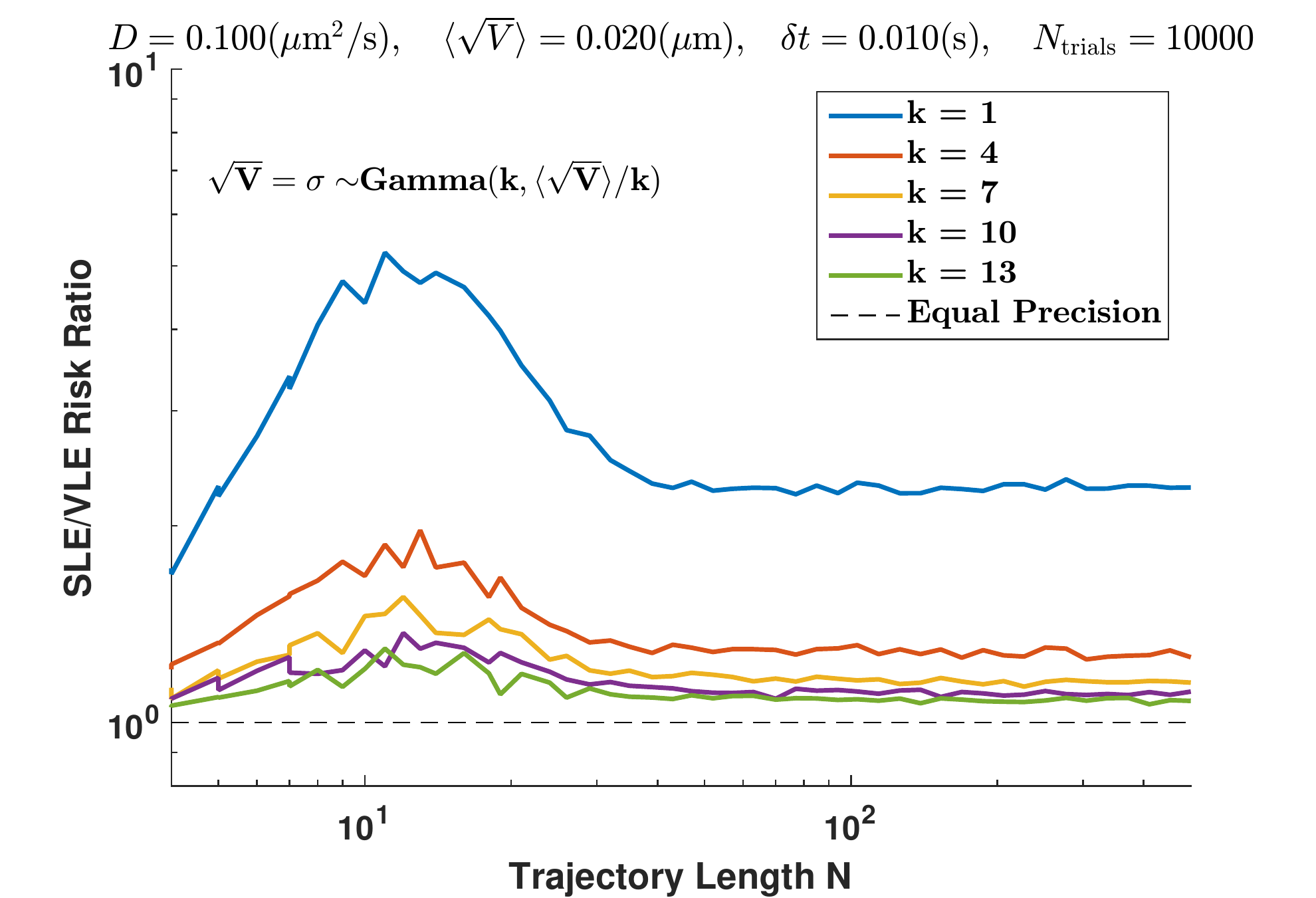} & \includegraphics[height=2.2in]{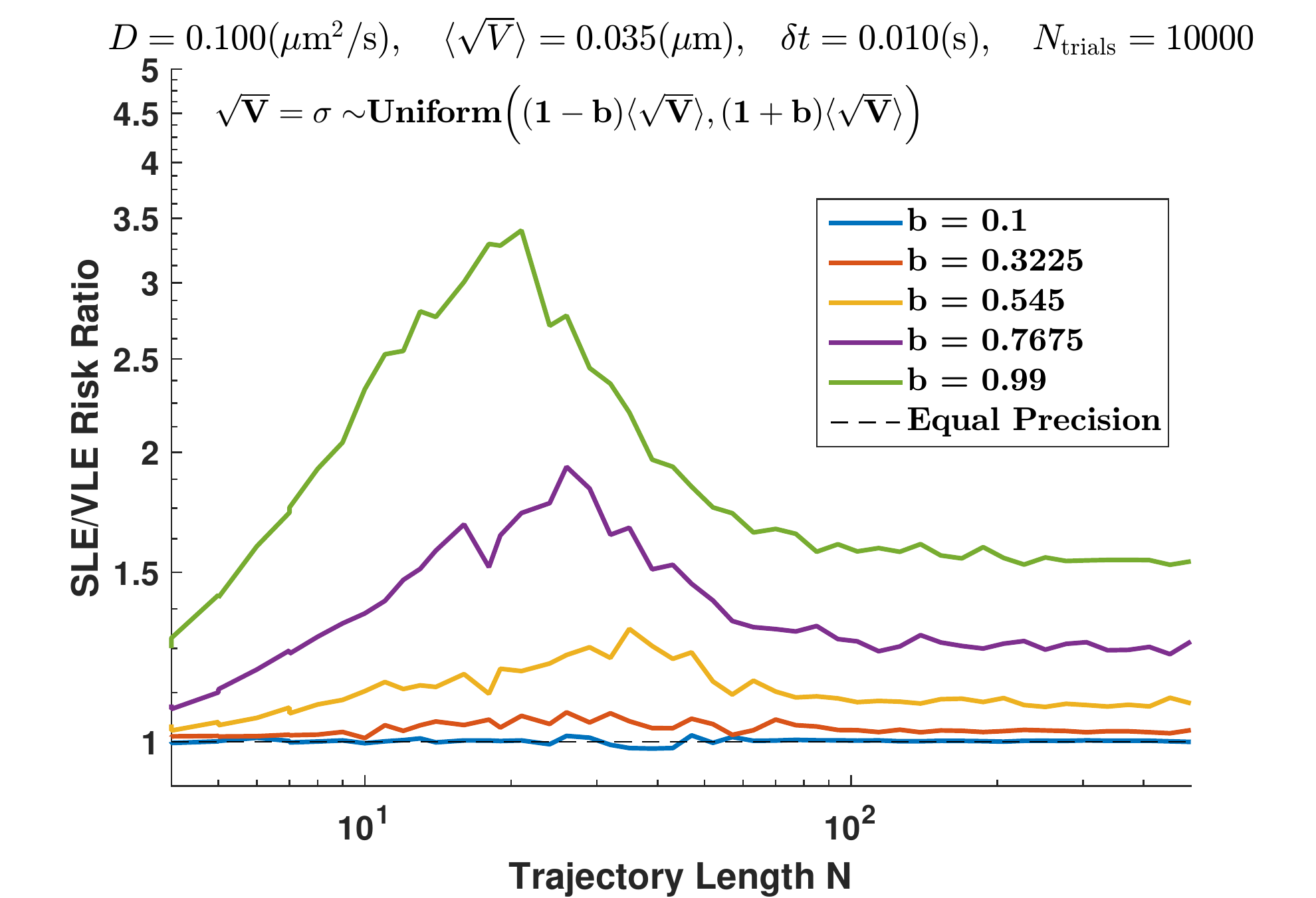}
\end{tabular}
\end{center}
\caption{Risk ratio of the Scalar Localization Error (SLE) and Vector of Localization Errors (VLE) MLEs for various trajectories parameterized by localization errors drawn from Gamma and Uniform distributions of a single varying parameter but the same mean $\langle \sqrt{V} \rangle$.  Log-log plot (A) shows the risk ratio of SLE and VLE with 5 Gamma distributions of the same mean localization error and different shape parameters, k.  The increased value of k reduces the variance of the gamma distribution; k = 1 is an exponential distribution and k = 13 is approaching a gaussian distribution.  All of the gamma distributions in (A) have a variance of $\langle \sqrt{V} \rangle^2 / k$.  Log-log plot (B) shows the risk ratio of SLE and VLE with 5 Uniform distributions of the same localization variance and different sampling boundaries, b.  The reduced value of b reduces the variance of the uniform distribution; a given b in plot (B) corresponds to a variance of $\left(\langle \sqrt{V} \rangle b\right)^2/3$, hence b = 0.1 is nearly a variance of 0.  As long as the variance of $\langle \sqrt{V} \rangle$ is greater than 0, the VLE estimators outperform the SLE estimators.}
\label{fig:varparam}
\end{figure}

We see in Fig.~\ref{fig:varparam} that the effect of increasing the variance relative to the mean of the localization errors in these test distributions results in a growing disparity between the two classes of estimators.  From Fig.~\ref{fig:risks} the effect of increasing $\langle \sqrt{V} \rangle$ sets a minimum trajectory length where estimates become reasonable; e.g the linear decrease in risk for the gamma distribution of Fig.~\ref{fig:risks}A is seen at shorter trajectories than the uniform distribution of Fig.~\ref{fig:risks}C even though both the gamma and uniform distributions have the same variance because the uniform distribution has a larger $\langle \sqrt{V} \rangle$.  In Fig.~\ref{fig:varparam} the constant precision improvement of the VLE estimators are recognized in the regime where both estimators start reporting risk values that scale linearly with trajectory length.  Prior to that, the VLE estimators start performing significantly better at shorter trajectories, which is represented by the peaks seen in Fig.~\ref{fig:varparam}. These simulations, while overly simplified versions of real SPT trajectories, highlight the importance of characterizing each localization error accordingly to their associated localizations in a trajectory.

\section{Discussion and Conclusion}

Starting from the fundamental diffusion likelihood expression (Eq.~\ref{eq:Funk}), we presented three independent solutions, each of which has different benefits, and leads to a computational algorithm with different advantages.  The recursive method presents the simplest solution that is numerically more stable than the other methods for estimating likelihoods when $D$ approaches 0.  The Laplace method has the advantage that the expected true positions $\hat{X}$ are computed along with the likelihood, which may be useful in some applications.  The Markov method was crucial for deriving the terms $\epsilon_i$ in the components $\M_i$ given knowledge of the true underlying probability distribution, hence the generality of the Markov method is its main advantage.  In terms of numerical accuracy and computational efficiency the Markov method is better than the Laplace method especially for very small $D$ values, but it remains computationally inferior to the recursive method. For practical implementations, we recommend the recursive method unless the MLE of the true positions is also desired.

The method described here naturally allows for trajectories with missing or irregularly spaced localizations by decoupling the concept of observation times $t_i$ from the exposure time $t_\epsilon$.  This is important when some localizations are missed because the gap between $t_i$ and $t_{i+1}$ becomes larger, but the exposure time remains the same.  Thus to correctly account for the motion-blur, the effective weighting of the dependence of observed position $o_i$ on $x_i$ and $x_{i+1}$ changes, and our technique directly incorporates this effect.  Trajectory intermittency has been accounted for in prior studies~\cite{Shuang2013} and in those same studies extensions to dynamic errors were suggested, but a convenient computational framework for an estimator that seamlessly factors in both trajectory intermittencies and dynamic error had not been explicitly worked out until now.

Numerical implementations of the likelihood forms resulting from the three derivations were tested to justify the equivalence among the likelihood forms.  Since all three derivations began from the same set of first principles, the three likelihood calculations are essentially equivalent.  We note however that our implementation remains unit agnostic, so that the trajectory with a very small $D$ value can be easily scaled up to appropriate units so that the value of $D$ is closer to 1, where the numerical calculations will be more robust.

Variable localization uncertainties could occur in practice from variable background intensities or photobleaching of the fluorescent label.  We compared the performance of our VLE estimator to the current state-of-the-art SLE estimator using the squared log loss function and found a clear performance benefit when trajectories had variable localization uncertainties.

\section{Author Contributions}
PKR and MJO contributed equally to this work. KAL, PJC and PKR conceived the project. PJC initiated the formulation of the estimation problem as a MLE given a set of observations. PKR derived the recursive, Markov, and Laplace methods. MJO derived the efficient algorithms for the three solution methods and the C++ and MATLAB implementations of the methods, generated the estimation accuracy results and helped to simplify the presentation. All authors contributed to the writing and editing of the manuscript.

\section{Acknowledgments}
We wish to acknowledge Stan Steinberg and Michael Wester for reading our manuscript and providing enlightening discussions and helpful comments.  Financial support for this work was provided primarily by the National Science Foundation grant 0954836. Additional support was provided by The New Mexico Spatiotemporal Modeling Center: NIH P50GM085273 (KAL), NIH grant 1R01GM100114 (KAL,MJO) and NIH grant 1R01NS071116 (KAL, MJO).

\bibliography{DiffusionTheory_bib}

\begin{thebibliography}{10}

\bibitem{Saxton1997a}
Michael~J Saxton and Ken Jacobson.
\newblock {SINGLE-PARTICLE TRACKING:Applications to Membrane Dynamics}.
\newblock {\em Annual Review of Biophysics and Biomolecular Structure},
  26:373--399, 1997.

\bibitem{Saxton2010}
Michael~J Saxton.
\newblock Two-dimensional continuum percolation threshold for diffusing
  particles of nonzero radius.
\newblock {\em Biophysical journal}, 99(5):1490--1499, 2010.

\bibitem{sanamrad2014single}
Arash Sanamrad, Fredrik Persson, Ebba~G Lundius, David Fange, Arvid~H
  Gynn{\aa}, and Johan Elf.
\newblock Single-particle tracking reveals that free ribosomal subunits are not
  excluded from the escherichia coli nucleoid.
\newblock {\em Proceedings of the National Academy of Sciences},
  111(31):11413--11418, 2014.

\bibitem{calderon2013quantifying}
Christopher~P Calderon, Michael~A Thompson, Jason~M Casolari, Randy~C
  Paffenroth, and WE~Moerner.
\newblock Quantifying transient 3d dynamical phenomena of single mrna particles
  in live yeast cell measurements.
\newblock {\em The Journal of Physical Chemistry B}, 117(49):15701--15713,
  2013.

\bibitem{monnier2012bayesian}
Nilah Monnier, Syuan-Ming Guo, Masashi Mori, Jun He, P{\'e}ter L{\'e}n{\'a}rt,
  and Mark Bathe.
\newblock Bayesian approach to msd-based analysis of particle motion in live
  cells.
\newblock {\em Biophysical journal}, 103(3):616--626, 2012.

\bibitem{jaqaman2008robust}
Khuloud Jaqaman, Dinah Loerke, Marcel Mettlen, Hirotaka Kuwata, Sergio
  Grinstein, Sandra~L Schmid, and Gaudenz Danuser.
\newblock Robust single-particle tracking in live-cell time-lapse sequences.
\newblock {\em Nature methods}, 5(8):695--702, 2008.

\bibitem{serge2008dynamic}
Arnauld Serg{\'e}, Nicolas Bertaux, Herv{\'e} Rigneault, and Didier Marguet.
\newblock Dynamic multiple-target tracing to probe spatiotemporal cartography
  of cell membranes.
\newblock {\em Nature Methods}, 5(8):687--694, 2008.

\bibitem{chenouard2014objective}
Nicolas Chenouard, Ihor Smal, Fabrice De~Chaumont, Martin Ma{\v{s}}ka, Ivo~F
  Sbalzarini, Yuanhao Gong, Janick Cardinale, Craig Carthel, Stefano Coraluppi,
  Mark Winter, et~al.
\newblock Objective comparison of particle tracking methods.
\newblock {\em Nature methods}, 2014.

\bibitem{mont2014new}
Alexander~D Mont, Christopher~P Calderon, and Aubrey~B Poore.
\newblock A new computational method for ambiguity assessment of solutions to
  assignment problems arising in target tracking.
\newblock In {\em SPIE Defense+ Security}, pages 90920J--90920J. International
  Society for Optics and Photonics, 2014.

\bibitem{yoon2008bayesian}
Ji~Won Yoon, Andreas Bruckbauer, William~J Fitzgerald, and David Klenerman.
\newblock Bayesian inference for improved single molecule fluorescence
  tracking.
\newblock {\em Biophysical journal}, 94(12):4932--4947, 2008.

\bibitem{Smith2010b}
Carlas~S Smith, Nikolai Joseph, Bernd Rieger, and Keith~a Lidke.
\newblock {Fast, single-molecule localization that achieves theoretically
  minimum uncertainty.}
\newblock {\em Nature methods}, 7(5):373--5, May 2010.

\bibitem{Qian1991a}
H~Qian, M~P Sheetz, and E~L Elson.
\newblock {Single particle tracking. Analysis of diffusion and flow in
  two-dimensional systems.}
\newblock {\em Biophysical journal}, 60(4):910--21, October 1991.

\bibitem{martin2002apparent}
Douglas~S Martin, Martin~B Forstner, and Josef~A K{\"a}s.
\newblock Apparent subdiffusion inherent to single particle tracking.
\newblock {\em Biophysical journal}, 83(4):2109--2117, 2002.

\bibitem{Savin2005}
Thierry Savin and Patrick~S Doyle.
\newblock {Static and dynamic errors in particle tracking microrheology.}
\newblock {\em Biophysical journal}, 88(1):623--38, January 2005.

\bibitem{Michalet2010}
Xavier Michalet.
\newblock {Mean square displacement analysis of single-particle trajectories
  with localization error: Brownian motion in an isotropic medium.}
\newblock {\em Phys Rev E Stat Nonlin Soft Matter Phys}, 82(4):041914, October
  2010.

\bibitem{Michalet2012}
Xavier Michalet and Andrew~J Berglund.
\newblock {Optimal diffusion coefficient estimation in single-particle
  tracking.}
\newblock {\em Physical review. E, Statistical, nonlinear, and soft matter
  physics}, 85(6 Pt 1):061916, June 2012.

\bibitem{Montiel2006}
D~Montiel, H~Cang, and H~Yang.
\newblock {Quantitative characterization of changes in dynamical behavior for
  single-particle tracking studies.}
\newblock {\em Journal of Physical Chemistry B}, 110:19763--70, 2006.

\bibitem{Berglund2010}
A~J Berglund.
\newblock {Statistics of camera-based single-particle tracking.}
\newblock {\em Phys Rev E Stat Nonlin Soft Matter Phys}, 82:011917, 2010.

\bibitem{Shuang2013}
Bo~Shuang, Chad~P Byers, Lydia Kisley, Lin-Yung Wang, Julia Zhao, Hiroyuki
  Morimura, Stephan Link, and Christy~F Landes.
\newblock {Improved analysis for determining diffusion coefficients from short,
  single-molecule trajectories with photoblinking.}
\newblock {\em Langmuir : the ACS journal of surfaces and colloids},
  29(1):228--34, January 2013.

\bibitem{vestergaard2014optimal}
Christian~L Vestergaard, Paul~C Blainey, and Henrik Flyvbjerg.
\newblock Optimal estimation of diffusion coefficients from single-particle
  trajectories.
\newblock {\em Physical Review E}, 89(2):022726, 2014.

\bibitem{de1774memoire}
PS~de~Laplace.
\newblock M{\'e}moire sur les suites r{\'e}curro-r{\'e}currentes et sur leurs
  usages dans la th{\'e}orie des hasards.
\newblock {\em M{\'e}m. Acad. Roy. Sci. Paris}, 6:353--371, 1774.

\bibitem{markov1912}
A.A. Markov.
\newblock {\em Wahrscheinlichkeitsrechnung}.
\newblock BG Teubner, 1912.

\bibitem{Chandrasekhar1943}
S.~Chandrasekhar.
\newblock Stochastic problems in physics and astronomy.
\newblock {\em Rev. Mod. Phys.}, 15:1--89, Jan 1943.

\bibitem{cahill2013physical}
Kevin Cahill.
\newblock {\em Physical Mathematics}.
\newblock Cambridge University Press, 2013.

\bibitem{Isserlis1918}
L.~Isserlis.
\newblock On a formula for the product-moment coefficient of any order of a
  normal frequency distribution in any number of variables.
\newblock {\em Biometrika}, 12(1/2):pp. 134--139, 1918.

\bibitem{el2004inverse}
Moawwad~EA El-Mikkawy.
\newblock On the inverse of a general tridiagonal matrix.
\newblock {\em Applied Mathematics and Computation}, 150(3):669--679, 2004.

\bibitem{brown1968inadmissibility}
L~Brown.
\newblock Inadmissibility of the usual estimators of scale parameters in
  problems with unknown location and scale parameters.
\newblock {\em The Annals of Mathematical Statistics}, pages 29--48, 1968.

\bibitem{gelman2014bayesian}
Andrew Gelman, John~B Carlin, Hal~S Stern, and Donald~B Rubin.
\newblock {\em Bayesian data analysis}, volume~2.
\newblock Taylor \& Francis, 2014.

\bibitem{ross1996stochastic}
Sheldon~M Ross et~al.
\newblock {\em Stochastic processes}, volume~2.
\newblock John Wiley \& Sons New York, 1996.

\bibitem{de1972elementary}
Conte de~Boor.
\newblock {\em Elementary numerical analysis}.
\newblock McGraw-Hill, 1972.

\bibitem{ElMikkawy:2004}
Moawwad~E.A. El-Mikkawy.
\newblock On the inverse of a general tridiagonal matrix.
\newblock {\em Applied Mathematics and Computation}, 150(3):669 -- 679, 2004.

\end{thebibliography}

\newpage

\begin{center}
{\Large SUPPORTING MATERIAL: Estimation of the Diffusion Constant from Intermittent Trajectories with
Variable Position Uncertainties}
\end{center}

\tableofcontents

\section{Implemented Notation and Gaussian Functions}
\label{sec:NormalIdent}
The derivations presented in the main text make heavy use of normalized Gaussian functions, which we denote as a function of three arguments,
\begin{equation*}
    \N(a,b,v) = \frac{1}{\sqrt{2 \pi v}} \exp{\left[-\frac{(a-b)^2}{2v}\right]} .
\end{equation*}
The Gaussian function defined this way is symmetric with respect to the first two position parameters, so that $\N(a,b,v)=\N(b,a,v)$, and the variance of the function is given by the third parameter.
Also, the normalization factor ensures that the Gaussian integrated over all space with respect to either of its position parameters is unity,
\begin{equation*}
\int^{\infty}_{-\infty} \d{a}\, \N(a,b,v)=\int^{\infty}_{-\infty} \d{b}\, \N(a,b,v)=1.
\end{equation*}
As a corollary, a normalized Gaussian has a useful scaling identity, for $q>0$,
\begin{equation*}
\N(a,b,v)=q\N(qa,qb,q^2 v).
\end{equation*}

Next, consider the case of the product of two normalized Gaussians sharing a common position parameter,
which can be rewritten as a product of two normalized Gaussians where the common parameter only appears in one of the two,
\begin{equation}
\label{eq:Gproduct2}
\begin{aligned}
\N(x,\mu_1,\eta_1)\N(x, \mu_2, \eta_2) &= \N( \mu_1, \mu_2, \eta_1+\eta_2) \N(x, \mu',\eta'),
\end{aligned}
\end{equation}
where,
\[ \mu'=\frac{\mu_1\eta_2+\mu_2\eta_1}{\eta_1+\eta_2}\quad \mbox{and,}\quad \eta'=\frac{\eta_1\eta_2}{\eta_1+\eta_2}.\]
Using Eq.~\ref{eq:Gproduct2}, the integral of the product of two
normalized Gaussians over a shared position parameter is itself a normalized
Gaussian in the other two (unintegrated) position parameters,
\begin{equation}
\label{eq:Gint2}
\begin{aligned}
\int \d{x} \N(x,\mu_1,\eta_1)\N(x,\mu_2,\eta_2) = \N( \mu_1, \mu_2, \eta_1+\eta_2) \int  \d{x} \N(x, \mu', \eta')
= \N( \mu_1, \mu_2, \eta_1+\eta_2).\\
\end{aligned}
\end{equation}
With two successive applications of Eq.~\ref{eq:Gproduct2}, the product of three Gaussians which share a common position parameter becomes,
\begin{equation}
\label{eq:Gproduct3}
\begin{aligned}
\N(x,\mu_1,\eta_1)\N(x, \mu_2, \eta_2)\N(x, \mu_3, \eta_3) &= \N( \mu_1, \mu_2, \eta_1+\eta_2) \N(x, \mu', \eta')\N(x, \mu_3, \eta_3),\\
 &= \N( \mu_1, \mu_2, \eta_1+\eta_2) \N(\mu_3, \mu',  \eta'+\eta_3) \N(x, \mu'', \eta''),\\
\mbox{where,}\quad \mu''&=\frac{\mu'\eta_3+\mu_3\eta'}{\eta'+\eta_3},\\
\mbox{and,}\quad \eta''&=\frac{\eta'\eta_3}{\eta'+\eta_3}.
\end{aligned}
\end{equation}
Finally, Eq.~\ref{eq:Gproduct3} allows the integral of three normalized Gaussians over a shared position parameter to reduce to
\begin{equation}
\label{eq:Gint3}
\begin{aligned}
\int \d{x} \N(x,\mu_1,\eta_1)\N(x,\mu_2,\eta_2)\N(x,\mu_3,\eta_3) &= \N( \mu_1, \mu_2, \eta_1+\eta_2) \N(\mu_3, \mu', \eta'+\eta_3) \int \d{x} \N(x, \mu'', \eta'')\\
&= \N( \mu_1, \mu_2, \eta_1+\eta_2) \N(\mu_3, \mu', \eta'+\eta_3)\\
&= \N( \mu_1, \mu_2, \eta_1+\eta_2) \N(\mu_3, \mu', \gamma),\\
\mbox{where,}\quad \gamma&= \frac{\eta_1\eta_2+\eta_1\eta_3+\eta_2\eta_3}{\eta_1+\eta_2}.
\end{aligned}
\end{equation}

\section{Problem Formulation}
\label{sec:ProbForm}

In the main manuscript, the likelihood distribution for a 1D random walk given a set of observations, $O=\{o_i\}_{i=1}^N$,
is described as an integral that marginalizes over the $N+1$ unknown true positions $X=\{x_i\}_{i=1}^{N+1}$,
\begin{equation} \label{eq:pod}
    \P{O}{D} = \intRNpone \d{X} \prod^{N}_{i=1} \P{o_i}{x_i,x_{i+1}} \P{x_{i+1}}{x_i}.
\end{equation}
We seek to rigorously define our formalism to justify the analysis used in the rest of the manuscript.

\subsection{The Probability of a Single Observation and the True Start Positions}

Under typical biological SPT experiments, the temporal resolution of a probe is high enough that the effects of diffraction are much greater than the effects of particle motion in generating the point spread function of the image.  Following the assumption that the variance due to diffraction is significantly greater than the variance due to motion, the point spread function can be defined as a stationary Gaussian function centered about the average position of the particle during a single frame with some background offset.   With these considerations, we now set to define $\P{o_i,x_i,x_{i+1}}{D}$: the probability of an observation and the particle's true start and end points in a frame given the free diffusion model.  Since the maximum likelihood estimator for a Gaussian function returns the peak (the maximum likelihood) and variance (the error) of a gaussian distribution, there is sufficient information from the estimator to build a probability distribution relating the localization to the true averaged position of a particle.  The probability of obtaining a localized position, $o_i$, given the true averaged position of the particle, $\bar{y}_i$, and the estimator variance, $v_i$, is
\begin{equation} \label{avtoobs}
    \P{o_i}{\bar{y}_i} = \N(o_i,\bar{y}_i,v_i).
\end{equation}
We incorporate more information on $\bar{y}_i$ by relating it to the start positions at the considered frame, $i$, and the subsequent frame, $i+1$; where it is assumed that a frame begins immediately after the prior frame ends.  Therefore the probability distribution of $o_i$ with frame start coordinates, $x_i$ and $x_{i+1}$, is expressed as
\begin{equation*}
\P{o_i,x_i,x_{i+1}}{D} = \int \d{\bar{y}_i} \; \P{o_i}{\bar{y}_i} \P{\bar{y}_i}{x_i,x_{i+1}} \P{x_{i+1}}{x_i} P(x_i),
\end{equation*}
where the diffusion constant, D, is implicitly included in the probability distributions.  The probabilities $\P{x_{i+1}}{x_i}$ represent a pure diffusion process.
\begin{equation} \label{pure_diff}
\P{x_{i+1}}{x_i} = \N(x_{i+1},x_i,\omega_i(D)),
\end{equation}
where $\omega_i(D)=2D\delta t_i$ and $\delta t_i = t_{i+1} - t_i$.

In section ~\ref{sec:ExpDeriv}, we will explicitly derive $\P{\bar{y}_i}{x_i,x_{i+1}}$, but here we will state the result as
\begin{equation*}
    \P{\bar{y}_i}{x_i,x_{i+1}} = \N\left(\bar{y}_i, \left(1 - \frac{t_{\epsilon}}{ 2 \delta t_i} \right) x_i + \left( \frac{t_{\epsilon}}{2 \delta t_i} \right) x_{i+1}, 2D t_{\epsilon} \left[\frac{1}{3} - \frac{t_{\epsilon}}{4 \delta t_i}  \right] \right),
\end{equation*}
where $t_\epsilon$ is the exposure time of a frame; in other words, the time the last photon observed in one frame can be arbitrarily spaced from the first photon in the next frame.  Hence, trajectory intermittencies can be accounted for by redefining $\delta t_i$ so that frame $i+1$ is the next frame that observes a photon from the particle under consideration, omitting all frames that do not provide measurement information. Given that $\P{\bar{y}_i}{x_i,x_{i+1}}$ is in the form of a Gaussian function with $\bar{y}_i$ as one of the location parameters, $\bar{y}_i$ can be effectively marginalized in $\P{o_i,x_i,x_{i+1}}{D}$ so that the expression reduces to
\begin{equation*}
    \P{o_i,x_i,x_{i+1}}{D} = \P{o_i}{x_i,x_{i+1}} \P{x_{i+1}}{x_i} P(x_i).
\end{equation*}
When obtaining the global probability of all $O$ given $D$, the probabilities $P(x_i)$ will be conditioned on prior observations, so it becomes necessary to define the expression
\begin{equation*}
    \P{o_i,x_{i+1}}{x_i,D} = \P{o_i}{x_i,x_{i+1}} \P{x_{i+1}}{x_i}.
\end{equation*}

\subsection{The Probability of a Trajectory of Observations}

We now wish to get the full expression for $\P{O}{D}$, which has no dependency on the true positions, $X$.  To do this, first define the expression
\begin{equation*}
\P{O,X}{D} = P(x_1) \prod^{N}_{i=1} \P{o_i,x_{i+1}}{x_i,D} = P(x_1) \prod^{N}_{i=1} \P{o_i}{x_i,x_{i+1}} \P{x_{i+1}}{x_i}.
\end{equation*}
Since $X$ are nuisance parameters, integrate the $X$ variables over all configuration space and set $P(x_1)=1$ because it is assumed that $x_1$ must already be known, since the inference of a diffusion probability must have an origin to relate all subsequent coordinates.
\begin{equation} \label{eq:pod2}
    \P{O}{D} = \intRNpone \d{X} \prod^{N}_{i=1} \P{o_i}{x_i,x_{i+1}} \P{x_{i+1}}{x_i}.
\end{equation}
We will see in section ~\ref{sec:ExpDeriv} that the following expression, which is the focus of the main manuscript, is equivalent and easier to evaluate
\begin{equation}
\label{eq:pod-normal-form}
    \P{O}{D} = \intRN \d{X} \prod^{N}_{i=1} \N(o_i,x_i,\eps_i(D)) \prod^{N-1}_{j=1} \N(x_{i+1},x_i,\omega_i(D)) = \intRN \d{X} \prod^{N}_{i=1} \M_i \prod^{N-1}_{j=1} \T_j.
\end{equation}
where
\begin{align}
    \label{eq:varM} \M_i = \M_i(o_i,x_i) = & \N(o_i,x_i,\eps_i(D)),\quad\textrm{for } 1\leq i\leq N, \textrm{ and } \\
    \label{eq:varT} \T_i = \T_i(x_{i+1},x_i) = & \N(x_{i+1},x_i,\omega_i(D)), \quad\textrm{for } 1\leq i \leq N-1.
\end{align}

\section{Explicit Derivations on the Problem Formulation}
\label{sec:ExpDeriv}

There were a few results presented in section ~\ref{sec:ProbForm} that are clarified in this section.  The probability distribution of an averaged position given the start and end points of a frame are discussed and analytically solved in the continuous limit.  Then the relation between representation of $\P{O}{D}$ with the probability distribution and reduced measurement functions is investigated.

\subsection{The Probability Density of a Time Averaged Position}
\label{sec:TimeAv}
Given $D$, the probability density of a transition from point $a$ to to point $b$ separated by a time $T$ is
\begin{equation*}
    \P{b}{a} = \N(b,a,2DT).
\end{equation*}
If an intermediate point, $y(t)$ sampled at a time $t<T$ is considered, the joint probability density of a transition from $a$ to $y(t)$ and then from $y(t)$ to $b$ is
\begin{align*}
    P(y(t),a,b) & = P(a) \N(y(t),a,2Dt) \N(b,y(t),2D(T-t)) \\
    & = P(a) \N(b,a,2DT) \N\left(y(t),a\left(1-\frac{t}{T}\right)+b\frac{t}{T},2D\frac{t}{T}(T-t)\right).
\end{align*}
The probability density of the variable $y(t)$, preconditioned on the end points, a and b, is then
\begin{equation} \label{bbridge}
    \P{y(t)}{b,a} = \N\left(y(t),a\left(1-\frac{t}{T}\right)+b\frac{t}{T},2D\frac{t}{T}(T-t)\right).
\end{equation}
Eq.~\ref{bbridge} is the probability density for what is known as a Brownian Bridge~\cite{ross1996stochastic}, or Brownian motion with preconditioned end points.  It has a mean and covariance defined as
\begin{align*}
    \E{y(t)} & = a\left(1-\frac{t}{T}\right) + b \frac{t}{T} \\
    \text{cov}[y(t),y(s)] & = 2D \left(s -\frac{st}{T}\right) \quad \text{for} \quad s\leq t \leq T
\end{align*}
It is now of interest to find the probability density for a quantity that describes an integrated average of $y(t)$ such that
\begin{equation*}
    \bar{y} = \frac{1}{t_{\eps}} \int^{t_{\eps}}_{t=0} \d{t} y(t).
\end{equation*}
Since each $y(t)$ is a normally distributed random variable, y(t) is a gaussian process, then the time averaged integral of a Gaussian process, $\bar{y}$, is also a normally distributed random variable.  Therefore, from Isserlis theorem~\cite{Isserlis1918} only the first two moments of $\bar{y}$ are needed to determine its probability distribution.  The first moment is
\begin{align*}
\E{\bar{y}} & = \E{\frac{1}{t_{\eps}} \int^{t_{\eps}}_{t=0} \d{t} y(t)} = \frac{1}{t_{\eps}} \int^{t_{\eps}}_{t=0} \d{t} \E{y(t)} \\
 & = \frac{1}{t_{\eps}} \int^{t_{\eps}}_{t=0} \d{t}  \left[ a\left(1-\frac{t}{T}\right) + b \frac{t}{T} \right] = a\left(1-\frac{t_{\epsilon}}{2T}\right) + b \frac{t_{\epsilon}}{2T} \\
 & = (1-\alpha)a + \alpha b,
\end{align*}
where for notational convenience, we define $\alpha = t_{\epsilon}/2T$.  It follows that the second moment is
\begin{align*}
\E{\bar{y}^2} & = \E{\frac{1}{t^2_{\eps}} \int^{t_{\eps}}_{t=0} \d{t} y(t) \int^{t_{\eps}}_{s=0} \d{s} y(s)} = \frac{1}{t^2_{\eps}} \int^{t_{\eps}}_{t=0} \d{t} \int^{t_{\eps}}_{s=0} \d{s} \E{ y(t) y(s)} \\
    & = \frac{1}{t^2_{\eps}} \int^{t_{\eps}}_{t=0} \d{t} \int^{t_{\eps}}_{s=0} \d{s} \E{ [y(t)-\E{y(t)}+\E{y(t)}] [y(s)-\E{y(s)} + \E{y(s)}]} \\
    & = \frac{1}{t^2_{\eps}} \int^{t_{\eps}}_{t=0} \d{t} \int^{t_{\eps}}_{s=0} \d{s} \text{cov}[y(t),y(s)] + \E{y(t)}\E{y(s)} \\
    & = \E{\bar{y}}^2 + \frac{1}{t^2_{\eps}} \int^{t_{\eps}}_{t=0} \d{t} \int^{t_{\eps}}_{s=0} \d{s} \text{cov}[y(t),y(s)].
\end{align*}
Therefore, the variance for $\bar{y}$ is
\begin{align*}
    \E{\bar{y}^2}-\E{\bar{y}}^2 & = \frac{1}{t^2_{\eps}} \int^{t_{\eps}}_{t=0} \d{t} \int^{t_{\eps}}_{s=0} \d{s} \text{cov}[y(t),y(s)] \nonumber \\
    & = \frac{1}{t^2_{\eps}} \int^{t_{\eps}}_{t=0} \d{t} \left[ \int^{t_{\eps}}_{s=t} \d{s} 2D \left( t - \frac{st}{T} \right) + \int^{t}_{s=0} \d{s} 2D \left(s -\frac{st}{T} \right) \right] \nonumber \\
    & = \frac{1}{t^2_{\eps}} \int^{t_{\eps}}_{t=0} \d{t} 2D \left[ t_{\epsilon}t - t^2 - \frac{t^2_{\epsilon}t}{2T} + \frac{t^2}{2} \right] \nonumber \\
    & = 2D \left[ \frac{t_{\epsilon}}{2} - \frac{t_{\epsilon}}{3} - \frac{t^2_{\epsilon}}{4T} + \frac{t_{\epsilon}}{6} \right] = 2D t_{\epsilon} \left[ \frac{1}{3} - \frac{\alpha}{2} \right].
\end{align*}
Given that $\bar{y}$ must be a normally distributed and its first two moments are known, then
\begin{equation*}
    \P{\bar{y}}{a,b} = \N(\bar{y},\E{\bar{y}},\E{\bar{y}^2}-\E{\bar{y}}^2) = \N\left(\bar{y},(1-\alpha) a+\alpha b,2Dt_{\eps}\left[\frac{1}{3}-\frac{\alpha}{2}\right]\right)
\end{equation*}
Furthermore, it was established in Eq.~\ref{avtoobs} that a time averaged position was related to a localized observation by a normal Gaussian function so that
\begin{equation} \label{eq:oprobcond}
    \P{o}{a,b} = \int \d{\bar{y}} \P{o}{\bar{y}} \P{\bar{y}}{a,b} =  \N\left(o,(1-\alpha)a+\alpha b,v + 2Dt_{\eps}\left[\frac{1}{3}-\frac{\alpha}{2}\right]\right)
\end{equation}

\subsection{Functional Form: From Products of Probability Components to Simpler Expressions}
\label{sec:simplerforms}

In the limit where the camera exposure time goes to 0, the probability of $o_i$ is dependent on one coordinate, $x_i$.  However, as the camera exposure time becomes non-negligible with respect to the time spacing between frames, the probability of $o_i$ becomes increasingly dependent on the subsequent coordinate, $x_{i+1}$.  The $o_i$ dependence on both $x_i$ and $x_{i+1}$ make a direct approach to solving the integral computationally difficult.  In the main manuscript, Markov's method approach showed the following relationship for a multivariate gaussian
\begin{equation*}
    \Sigma_{i,j} = \E{\jump_i \jump_j},
\end{equation*}
where $\Sigma$ is the covariance matrix of a multivariate Gaussian function describing the vector of random variables $\Jump=\{\jump_i = o_{i+1}-o_i\}_{i=1}^{N-1}$.  Given that our probability distribution is obtained by integrating several Gaussian functions, the result of the distribution is a Gaussian function.  Therefore, if the moments of $\Jump$ are known, the parameter $\epsilon_i$ for the simpler expression can be derived given that
\begin{equation*}
    \E{\jump_i \jump_{i+1}} = - \epsilon_{i+1}
\end{equation*}
was previously shown to be true for the simpler expression in the main manuscript.  Starting from our derived probability expression in Eq.~\ref{eq:oprobcond} for an arbitrary $o_i$
\begin{equation*}
    \P{o_i}{x_i,x_{i+1}} = \N\left(o_i, (1 - \alpha_i ) x_i + \alpha_i x_{i+1}, v_i + 2D t_{\epsilon} \left[\frac{1}{3} - \frac{\alpha_i}{2}  \right] \right)
\end{equation*}
where $v_i$ is defined as the localization variance and
\begin{equation*}
\alpha_i = \frac{t_{\epsilon}}{2 \delta t_i}.
\end{equation*}
From the properties of a Gaussian function with 0 mean
\begin{align} \label{pro_exp}
    \E{o_i - (1 - \alpha_i ) x_i - \alpha_i x_{i+1}} & = 0, \nonumber \\
    \E{(o_i - (1 - \alpha_i ) x_i - \alpha_i x_{i+1})^2} & = v_i + 2D t_{\epsilon} \left[\frac{1}{3} - \frac{\alpha_i}{2}  \right].
\end{align}
Which implies
\begin{align} \label{avexpvarm}
    \E{o_i-x_i} & = 0 \nonumber \\
    \E{o_i-x_i|X} & = \alpha_i(x_{i+1}-x_i) \nonumber \\
    \E{(o_i-x_i)^2} & = v_i + 2Dt_{\epsilon} \frac{1}{3}.
\end{align}
Also
\begin{equation} \label{oiexp}
    \E{\jump_i} = \E{o_{i+1}-o_i} = \E{o_{i+1}-x_{i+1}} - \E{o_i-x_i} + \E{x_{i+1}-x_i} = 0.
\end{equation}
Given the relations in Eq.~\ref{avexpvarm} and Eq.~\ref{oiexp}
\begin{align*}
    \E{\jump_i \jump_{i+1}} & = \E{(o_{i+2}-o_{i+1})(o_{i+1}-o_i)} = - \epsilon_{i+1} \\
    & = \E{(o_{i+2}-o_{i+1}+x_{i+2}-x_{i+2}+x_{i+1}-x_{i+1})(o_{i+1}-o_i+x_{i+1}-x_{i+1}+x_i-x_i)} \\
    & = \E{(o_{i+1}-x_i)(x_{i+1}-x_i)} - \E{(o_{i+1}-x_{i+1})^2} + \cdots \\
    & = \alpha_{i+1}(2D \delta t_{i+1}) -2D t_{\epsilon} \frac{1}{3} - v_{i+1} = 2D t_{\epsilon} \frac{1}{6} - v_{i+1}.
\end{align*}
Where all the other terms in the ellipsis ($\cdots$) go to 0.  Additionally, it follows that
\begin{equation*}
    \E{\jump_i^2} = w_i + \epsilon_i + \epsilon_{i+1} = 2D \delta t_i - 4D t_{\epsilon} \frac{1}{6} + v_i + v_{i+1}.
\end{equation*}
Therefore
\begin{equation*}
    \eps_i(D) = v_i - 2D t_{\epsilon} \frac{1}{6}.
\end{equation*}
Which is the variance correction discovered in earlier diffusion estimation papers \cite{Savin2005,Montiel2006,Berglund2010}.

\section{Method Component Derivations}
\label{sec:MethCom}
The following sub-sections explain some of the relations that were explicitly stated to complete the derivations in the main text.

\subsection{Laplace Method: Maximum Likelihood of True Positions}

Recalling the objective function in the Laplace method
\begin{equation} \label{loggausint}
    -\text{ln}(f(\textbf{X})) = \sum^N_{i=1} \left[ \frac{1}{2}\text{ln}(2\pi\eps_i) + \frac{(o_i-x_i)^2}{2\eps_i} \right] + \sum^{N-1}_{i=1} \left[ \frac{1}{2}\text{ln}(2\pi\omega_i) + \frac{(x_{i+1}-x_i)^2}{2\omega_i} \right],
\end{equation}
the gradient of Eq.~\ref{loggausint} is
\begin{align} \label{laJacobian}
    -\frac{\partial \text{ln}f}{\partial x_1} & = \frac{(x_1-o_1)}{\eps_1} + \frac{(x_1-x_2)}{\omega_1} \nonumber \\
    -\frac{\partial \text{ln}f}{\partial x_i} & = \frac{(x_i-o_i)}{\eps_i} +\frac{(x_i-x_{i-1})}{\omega_{i-1}}  +\frac{(x_i-x_{i+1})}{\omega_i} \nonumber \\
    -\frac{\partial \text{ln}f}{\partial x_N} & = \frac{(x_N-o_N)}{\eps_N}  +\frac{(x_N-x_{N-1})}{\omega_{N-1}},
\end{align}
 where $i \in 2:N-1$. The Hessian $-\text{ln} \nabla \nabla f(\widehat{X}) = M$, of Eq.~\ref{loggausint} has the non-zero elements
\begin{align} \label{LaHessian}
    M_{1,1} & = -\frac{\partial^2\text{ln}f}{\partial x_1\partial x_1} = \frac{1}{\eps_1}+\frac{1}{\omega_1} \nonumber \\
    M_{i,i} & = -\frac{\partial^2\text{ln}f}{\partial x_i\partial x_i} = \frac{1}{\eps_i}+ \frac{1}{\omega_{i-1}} +\frac{1}{\omega_i}  \nonumber \\
    M_{N,N} & = -\frac{\partial^2\text{ln}f}{\partial x_N\partial x_N} = \frac{1}{\eps_N}+\frac{1}{\omega_{N-1}} \nonumber \\
    M_{i,i+1} & = -\frac{\partial^2\text{ln}f}{\partial x_i\partial x_{i+1}} = -\frac{1}{\omega_i} \nonumber \\
    M_{i,i-1} & = -\frac{\partial^2\text{ln}f}{\partial x_i\partial x_{i-1}} = -\frac{1}{\omega_{i-1}} \nonumber
\end{align}
Setting the gradient in Eq.~\ref{laJacobian} equal to 0 and moving the constants to the left hand side of the equation gives
\begin{align*}
    \frac{o_1}{\eps_1} & = \frac{\widehat{x}_1}{\eps_1} + \frac{(\widehat{x}_1-\widehat{x}_2)}{\omega_1} \\
    \frac{o_i}{\eps_i} & = \frac{\widehat{x}_i}{\eps_i} +\frac{(\widehat{x}_i-\widehat{x}_{i-1})}{\omega_{i-1}} +\frac{(\widehat{x}_i-\widehat{x}_{i+1})}{\omega_i} \\
    \frac{o_N}{\eps_N} & = \frac{\widehat{x}_N}{\eps_N}  +\frac{(\widehat{x}_N-\widehat{x}_{N-1})}{\omega_{N-1}}.
\end{align*}
With additional factoring, the expression looks like
\begin{align}
    \frac{o_1}{\eps_1} & = \widehat{x}_1\cdot\left(\frac{1}{\eps_1} + \frac{1}{\omega_1}\right) + \widehat{x}_2\cdot\left(\frac{-1}{\omega_1}\right) \nonumber \\
    \frac{o_i}{\eps_i} & = \widehat{x}_i\cdot\left(\frac{1}{\eps_i} + \frac{1}{\omega_{i-1}} + \frac{1}{\omega_i}\right) +\widehat{x}_{i-1}\cdot\left(\frac{-1}{\omega_{i-1}}\right)  +\widehat{x}_{i+1}\cdot\left(\frac{-1}{\omega_i}\right) \nonumber \\
    \frac{o_N}{\eps_N} & = \widehat{x}_N\cdot\left(\frac{1}{\eps_N} +\frac{1}{\omega_{N-1}}\right)  +\widehat{x}_{N-1}\cdot\left(\frac{-1}{\omega_{N-1}}\right). \nonumber
\end{align}

The factored expression on the right can be expressed in terms of a vector product of the Hessian matrix and the maximum likelihood of the true positions, $M \cdot \widehat{X}$. We invert the Hessian matrix to bring it to the other side of the equation so that the resulting expression for the maximum likelihood looks like
\begin{equation*} \label{MLE_X}
    \widehat{X} = M^{-1} \Theta,
\end{equation*}
where the components of $\Theta$ are
\begin{equation*} \label{theta_eval}
\theta_i = o_i/\eps_i.
\end{equation*}

\subsection{Laplace Method: Direct Integration of the Probability Distribution Components}

Starting from the probability component formalism

\begin{equation*}
    f(X) = \prod^{N}_{i=1} \P{o_i}{x_i,x_{i+1}} \P{x_i}{x_{i+1}} = \prod^{N}_{i=1} \N(o_i,(1-\alpha_i) x_i + \alpha_i x_{i+1}, q_i) \N(x_i,x_{i+1},\omega_i),
\end{equation*}
where $q_i$ is the variance due to the observation and $\alpha_i = \frac{t_{\epsilon}}{2 \delta t_i}$. We solve for the Hessian of our objective function $M = - \nabla \nabla \text{ln} f(X)$
\begin{align*}
    M_{1,1}   &= -\frac{\partial^2\ln f}{\partial x_1^2} = \frac{(1-\alpha_1)^2}{q_1}+\frac{1}{\omega_1}  \\
    M_{i,i}   &= -\frac{\partial^2\ln f}{\partial x_i^2} = \frac{(1-\alpha_i)^2}{q_i}+ \frac{(\alpha_{i-1})^2}{\eps_{i-1}}+ \frac{1}{\omega_i}+\frac{1}{\omega_{i-1}}, \quad 2\leq i \leq N  \\
    M_{N+1,N+1}   &= -\frac{\partial^2\ln f}{\partial x_N^2} = \frac{\alpha_N^2}{q_N}+\frac{1}{\omega_N}  \\
    M_{i,i+1} = M_{i+1,i} &= -\frac{\partial^2\ln f}{\partial x_i\partial x_{i+1}} = \frac{(1-\alpha_i)(\alpha_i)}{q_i} - \frac{1}{\omega_i}, \quad 1\leq i \leq N.
\end{align*}
The form is similar to our functions, so we can solve for the maximum likelihood in the same fashion
\begin{equation}
    \widehat{X} = M^{-1} \Theta \nonumber
\end{equation}
Where the components of $\Theta$ are modified as
\begin{equation}
\theta_i = \frac{(1-\alpha_i) o_i}{q_i} + \frac{\alpha_{i-1} o_{i-1}}{q_{i-1}}, \nonumber
\end{equation}
and for completeness we set $\alpha_0 = 0$ and $\alpha_{N+1} = 0$.

\subsection{Markov Method: Marginalizing the Likelihood Function}

To understand how the integration of $\Jump'$ on $\P{O}{D}$ behaves, lets first consider the integration of $O$ on $\P{O}{D}$ with one $o_i$ held constant, which is equivalent to multiplying $\P{O}{D}$ with a delta distribution $\delta(o_i'-o_i)$.  The integral of $\P{O}{D}$ and the delta distribution with respect to $O$ is of the form
\begin{align}
    & \int \d{O} \; \delta(o_i'-o_i) P(O|D) =  \int \d{O} \d{X} \; \delta(o_i'-o_i) \prod^N_{i=1} \M_i \prod^{N-1}_{j=1} \T_j \nonumber \\
    & = \int \d{X} \d{O} \; \delta(o_i'-o_i) \prod^N_{i=1} \N(o_i,x_i,\eps_i) \prod^{N-1}_{j=1} \N(x_{j+1},x_j,\omega_j). \nonumber
\end{align}
Shuffling the order of integration, so that the $O$ basis is integrated first allows us to marginalize all $O$ except for $o_i$.  There are then N terms of $X$ which can be effectively marginalized
\begin{align}
    & \int \d{X} \d{O} \; \delta(o_i'-o_i) \prod^N_{i=1} \N(o_i,x_i,\eps_i) \prod^{N-1}_{j=1} \N(x_{j+1},x_j,\omega_j) \nonumber \\
    & = \int \d{X} \N(o_i',x_i,\eps_i) \prod^{N-1}_{j=1} \N(x_{j+1},x_j,\omega_j) = 1. \nonumber
\end{align}

Now we wish to perform a similar integration, but this time on the basis of $\Jump'$.  The integration of $\P{O}{D}$ with respect to $O$ is a completely different function than an integration with respect to $\Jump'$, but we wish to show that the integration on $\Jump'$ yields analogous to results to integration on $O$ with one $o_i$ held constant.  If $o_i=o_i'$ is held constant, than we can directly express $\jump'_i = o_{i+1} - o_i'$ in terms of one variable, $o_{i+1}$, if $i<N$.  We can also express $\jump'_{i-1}$ in terms of $o_{i-1}$ if $i>1$.  Analogously, every $o_{i + k}$ can be expressed as $$o_{i+k} = \sum^{i+k-1}_{j=i} \jump'_j + o_i' = \jump'_{i+k-1} + o_i' + g(\jump'_{i+k-2},i)$$
and every $o_{i - l}$ can be expressed as
$$o_{i-l} = \sum^{i-1}_{j=i-l} -\jump'_j + o_i' = -\jump'_{i-l} + o_i' + h(\jump'_{i-l+1},i).$$
We perform this substitution with an arbitrary $o_i$ held fixed to express the integral over $\Jump'$ as
\begin{align}
    & \int \d{\Jump'} \; P(O|D) =  \int \d{\Jump'} \d{X} \; \prod^N_{i=1} \M_i \prod^{N-1}_{j=1} \T_j \nonumber \\
    = & \int \d{X} \d{\Jump'} \; N(o_i',x_i,\eps_i) N(\jump_i' + o_i', x_{i+1}, \eps_{i+1}) \prod^{N}_{j=i+2} \N(\jump_{j-1}' + o_i' + g(\jump_{j-2},i),x_i,\eps_i) \nonumber \\
    & \prod^{i-1}_{k=1} \N(-\jump'_1 + o_i' + h(\jump'_2,i),x_1,\eps_1) \prod^{N-1}_{l=1} \N(x_{l+1},x_l,\omega_l). \nonumber
\end{align}
We can then shuffle the order of integration so that we can iteratively integrate the components of the $\Jump$ basis, essentially the components furthest from $o_i'$, that are expressed in only one of the univariate gaussian functions that comprise $\P{O}{D}$, effectively performing a sequential marginalization
\begin{align} \label{phitrick}
    & \int \d{X} \d{\Jump'} \; N(o_i',x_i,\eps_i) N(\jump_i' + o_i', x_{i+1}, \eps_{i+1}) \prod^{N}_{j=i+2} \N(\jump_{j-1}' + o_i' + g(\jump_{j-2},i),x_i,\eps_i) \nonumber \\
    & \prod^{i-1}_{k=1} \N(-\jump'_1 + o_i' + h(\jump'_2,i),x_1,\eps_1) \prod^{N-1}_{l=1} \N(x_{l+1},x_l,\omega_l) \nonumber \\
    = & \int \d{X} \d{\jump_i} \d{\jump_{i-1}} \N(o_i',x_i,\eps_i) N(\jump_i' + o_i', x_{i+1}, \eps_{i+1}) \N(-\jump_{i-1}' + o_i',x_i,\eps_i) \prod^{N-1}_{l=1} \N(x_{l+1},x_l,\omega_l) \nonumber \\
    = & \int \d{X} \N(o_i',x_i,\eps_i) \prod^{N-1}_{l=1} \N(x_{l+1},x_l,\omega_l)= 1.
\end{align}
From this result, we see that holding a single $o_i$ fixed is quite arbitrary, as the term is eventually marginalized by its associated $x_i$.  It is also apparent that fixing a particular $o_i$ allows complete isolation of a particular $\jump_i$ basis if all other $\jump_k$ bases are marginalized.  However, if we wish to evaluate $\jump_i$ and $\jump_{i-1}$ components with $o_i$ fixed, we see in the integral expression that there will be some correlation between adjacent displacements.  Most importantly, we see that $\P{O}{D}$ is a normalized probability density under $\Jump'$.

\subsection{Markov Method: Expectation Calculations}

We shall solve for the expectation values on $\Jump'$ over $\P{O}{D}$.  In order to do so, we will use the trick in Eq.\ref{phitrick} where we hold one observation constant until the appropriate substitutions are taken.  Solving for $\langle \jump_i \rangle$ in this spirit yields

\begin{align} \label{meanexpt}
    \langle \jump_i \rangle = & \int \d{\Jump'} \d{X} \jump_i  \prod^N_{j=1} \M_j \prod^{N-1}_{k=1} \T_k \nonumber \\
    = & \int \d{X} \d{\jump'_i} \; \jump'_i N(\jump_i' + o_i, x_{i+1}, \eps_{i+1}) \N(o_i,x_i,\eps_i) \prod^{N-1}_{j=1} \N(x_{j+1},x_j,\omega_j) \nonumber \\
    = & \int \d{X} \d{\jump'_i} \; \jump'_i N(\jump_i', 0, \eps_i + \eps_{i+1} + \omega_i) \N(x_i,\beta_i,\gamma_i) \N(x_{i+1},\beta_{i+1},\gamma_{i+1}) \prod^{N-1}_{j=1,j \neq i} \N(x_{j+1},x_j,\omega_j) \nonumber \\
    = & \int \d{\jump'_i} \; \jump'_i N(\jump_i', 0, \eps_i + \eps_{i+1} + \omega_i) = 0,
\end{align}
where the terms $\beta_i$ and $\gamma_i$ are intermediate variables that are generated from rearranging Gaussian functions.  These variables are immediately marginalized by integration on X, so we shall omit their explicit representation.  Aside from the fact that the expectation value on the displacement means are all 0, we also discover another striking property from Eq.~\ref{meanexpt}, that is, the marginalization of all other $\jump'$ and all $X$ isolates the expectation calculation to components of $\jump'_i$ that are independent of all other marginalized variables.  The same results from Eq.~\ref{meanexpt} are used so that the expectation of the variance is

\begin{align}
    \langle \jump_i^2 \rangle = & \int \d{\Jump'} \d{X} \; \jump^{'2}_i \prod^N_{j=1} \M_j \prod^{N-1}_{k=1} \T_k \nonumber \\
    = & \int \d{X} \d{\jump'_i} \; \jump^{'2}_i N(\jump_i' + o_i, x_{i+1}, \eps_{i+1}) \N(o_i,x_i,\eps_i) \prod^{N-1}_{j=1} \N(x_{j+1},x_j,\omega_j) \nonumber \\
    = & \int \d{X} \d{\jump'_i} \; \jump^{'2}_i N(\jump_i', 0, \eps_i + \eps_{i+1} + \omega_i) \N(x_i,\beta_i,\gamma_i) \N(x_{i+1},\beta_{i+1},\gamma_{i+1}) \prod^{N-1}_{j=1,j \neq i} \N(x_{j=1},x_j,\omega_j) \nonumber \\
    = & \int \d{\jump'_i} \; \jump^{'2}_i N(\jump_i', 0, \eps_i + \eps_{i+1} + \omega_i) =  \eps_i + \eps_{i+1} + \omega_i. \nonumber
\end{align}

Now on to the covariance terms; starting with adjacent displacements $\langle \jump_i \jump_{i+1} \rangle$.  However, if a given $o_i$ is fixed, there is a dependence between adjacent displacements.  With the techniques implemented in Eq.~\ref{meanexpt} a bivariate gaussian function is isolated from the rest of the integral expression.
\begin{align}
    \langle \jump_i' \jump_{i+1}' \rangle = & \int \d{\Jump'} \d{X} \; \jump_i' \jump_{i+1}' \prod^N_{j=1} \M_j \prod^{N-1}_{k=1} \T_k \nonumber \\
    = & \int \d{X} \d{\jump_i'} \d{\jump_{i-1}'} \; \jump_i' \jump_{i+1}' N(\jump_i' + o_i, x_{i+1}, \eps_{i+1}) \N(o_i,x_i,\eps_i) \N(\jump_{i-1}' + o_i,x_i,\eps_i) \prod^{N-1}_{j=1} \N(x_{j+1},x_j,\omega_j) \nonumber \\
    = & \int \d{\jump'_i} \d{\jump'_{i+1}} \; \jump_i' \jump_{i+1}' \N(S_b,0,\Sigma_b) = -\eps_{i+1}. \nonumber
\end{align}
Where the substitution of variables required for integrating the expectation of $\E{\jump_i' \jump_{i+1}'}$ induces a bivariate Gaussian function with location parameters $S_b = [s_i, s_{i+1}]^\transp$ and covariance matrix
\begin{equation}
    \Sigma_b = \begin{bmatrix} \omega_i + \epsilon_i + \epsilon_{i+1} & -\epsilon_{i+1} \\
        -\epsilon_{i+1} & \omega_{i+1} + \epsilon_{i+1} + \epsilon_{i+2} \end{bmatrix}. \nonumber
\end{equation}
Conversely, when evaluating $\langle \jump_i \jump_{i+k} \rangle$ where $k>1$, it is necessary to temporarily fix $o_i$ as well as $o_{i+k}$, this results in isolating two univariate Gaussian functions which results in
\begin{align}
    \langle \jump_i' \jump_{i+k}' \rangle = & \int \d{\Jump'} \d{X} \; \jump_i' \jump_{i+k}' \prod^N_{j=1} \M_j \prod^{N-1}_{k=1} \T_k \nonumber \\
      = & \int \d{\jump'_i} \d{\jump'_{i+k}} \; \jump_i' \jump_{i+k}'  N(\jump_i', 0, \eps_i + \eps_{i+1} + \omega_i) N(\jump_{i+k}', 0, \eps_{i+k} + \eps_{i+k+1} + \omega_{i+k})= 0. \nonumber
\end{align}

\section{Implementation}
\label{sec:Imp}

\subsection{Log--product Computation}
\label{sec:logprod-comp}
A common algorithmic problem shared by each of the three methods is the need to compute
logarithms of products.  A straight forward method is to utilize the identity
\begin{equation}
\label{eq:logprod}
\ln\left(\prod_{i=1}^N a_i \right)=\sum_{i=1}^N \ln a_i,
\end{equation}
the left side of which requires one logarithm and $N-1$ multiplications, while the right side requires $N$ logarithms and $N-1$ additions.
Directly using the left side of Eq.~\ref{eq:logprod} in computations can lead to numerical over- or underflow, while the $N$ logarithms required for the right side can dominate the computational costs, taking up the majority of the computational cycles for each of the three algorithms.  Thus, to minimize the number of logarithm computations, yet still maintain numerical accuracy, we utilize a hybrid log-product implementation to evaluate forms like Eq.~\ref{eq:logprod}.  The log--product method builds up a product of $a_i$-values, only taking a logarithm when multiplying by the next $a_i$ would lead to loss of precision, overflow, or underflow.

\subsection{Computation of Variance terms}
\label{sec:variance-comp}
Given the inputs, each of the method firsts compute the $N$ variance terms due to measurement, $\eps_i(D) = v_i - D t_{\epsilon}/3$, and the $N-1$ variance terms due to diffusion, $\omega_i(D) = 2D \delta t_i$.  Any of these variance terms can be arbitrarily close to zero, so to prevent numerical instabilities we bound these terms away from zero by at least machine epsilon, while preserving the sign which can be negative for $\eps_i(D)$.

\subsection{The Recursive Method Algorithm}
\begin{lstlisting}[title={}, caption={The Recursive Algorithm for the log-likelihood calculation in C++.  This fundemental algorithm can be impemented with just the log function from the C++ standard math library.  We rely on the variances to be computed as described in Sec.~\ref{sec:variance-comp}, and the log--product computation as described in Sec.~\ref{sec:logprod-comp}}]
FloatT recusiveLLH(int N,
                   const FloatT Obs[],
                   const FloatT dT[],
                   const FloatT vD[],
                   const FloatT vM[])
{
    FloatT alpha[N-1];
    FloatT eta = vD[0]+vM[0];
    FloatT mu = Obs[0];
    FloatT LLH = 0;
    for(int n=1;n<N-1;n++){
        FloatT temp_alpha = vM[n]+eta;
        alpha[n-1] = temp_alpha;
        FloatT temp_diff = (Obs[n]-mu);
        LLH += temp_diff*temp_diff/temp_alpha;
        mu = (mu*vM[n]+Obs[n]*eta)/temp_alpha;
        eta = vM[n]*eta/temp_alpha+vD[n];
    }
    alpha[N-2] = vM[N-1]+eta;
    LLH += (N-1)*log2pi;
    LLH += logprod(alpha);
    FloatT temp_diff = (Obs[N-1]-mu);
    LLH += temp_diff*temp_diff/alpha[N-2];
    LLH *= -0.5;
    return LLH;
}
\end{lstlisting}

\subsection{Tri-Diagonal matrix algorithms}
The Laplace and Markov method both require solving linear systems of the form $Ax=b$, where $A$ is symmetric tri-diagonal (all non-zero terms are on the main diagonal or those diagonals immediately above and below the main diagonal).
A naive solution based on inverting matrix $A$ has computational complexity $\bigO{N^3}$, where $N$ is the length of
the trajectory.  However because $A$ is tri-diagonal there are established algorithms~\cite{de1972elementary} that use the Gaussian elimination strategy to solve the system in time $\bigO{N}$.  Furthermore, the determinant of the matrix can
be shown to follow a recurrence relation~\cite{ElMikkawy:2004} that leads to a linear time computation.  Combined with the log--product computation described in Sec.~\ref{sec:logprod-comp}, this leads to a fast algorithm for computing $\log(\det(A))$ with a minimum number of calls to the logarithm function, and time complexity $\bigO{N}$.  We make use of these algorithms for the Laplace and Markov method implementations.

\subsection{DST Algorithm}
The DST algorithm code is based on the Matlab code provided by the authors~\cite{Michalet2012}.  Our implementation is as faithful as possible
to the original implementation but there are a few caveats that should be mentioned.  First because we assume that the localization variances are known, we use the form of the estimator that takes in the mean localization variance as an input parameter.  Because the DST can only incorporate a single localization
variance we use the mean of the given localization variances.  For the $R$ parameter we use $t_\epsilon/6$ as was derived for uniform exposure intervals~\cite{Michalet2012}.  The original implementation also leaves off the constant term $-(1/2) (N-1) \log(2 \pi)$ from the likelihood calculation.  While for MLE estimation this is not important, it does become important for other downstream analysis using the likelihood, and our algorithms do include this term, so we have added that in to the DST to make the plotted comparisons more fair.  Also, we ensure that the resulting values are always real as the original code
can return complex floating point numbers, but only the real parts have meaning.  Finally, because of the reliance on the discrete sine transform function which is
provided by Matlab, but not available natively in the C++ standard math library, the speed results from the main paper only test the provided Matlab implementation.

\end{document}